\documentclass[twocolumn]{aastex631}
\usepackage{natbib, color, subfigure,amsmath,soul}
\newcommand{\hbeta}{H{$\beta$}\,}
\newcommand{\halpha}{H{$\alpha$}\,}
\newcommand{\lya}{Ly\,$\alpha$\,}
\newcommand{\CIII}{C\,{\tt III]}\,}

\newcommand{\MgII}{[Mg\,{\tt II}]\,}
\def\CIV{C\,{\tt IV}\,}
\def\MgII{Mg\,{\tt II}\,}
\def\FeII{Fe\,{\tt II}\,}

\newcommand{\HeIIUV}{He\,{\tt II}\,$\lambda$1640\,}
\newcommand{\HeIIopt}{He\,{\tt II}\,$\lambda$4687\,}
\def\SiIV{Si\,{\tt IV}\,}

\def\L5100{\textit{$\lambda L_{5100}$}}
\def\kms{\textit{$\rm km\,s^{-1}$}}
\def\keV{\textit{$\rm keV$}}
\def\ergs{\textit{$\rm erg\ s^{-1}$}}
\def\Mbh{\textit{$M_{\rm BH}$}}

\def\dotm{\textit{$\dot{m}$}}
\def\Redd{\textit{$\lambda_{\rm Edd}$}}
\shorttitle{BLR photoionization}
\shortauthors{Wu et~al.}
\graphicspath{./} 

\begin{document}

\title{Understanding the Broad-line Region of Active Galactic Nuclei with Photoionization. I. the Moderate-Accretion Regime} 

\author[0000-0003-4202-1232]{Qiaoya Wu}
\email{qiaoyaw2@illinois.edu}
\affiliation{Department of Astronomy, University of Illinois at Urbana-Champaign, Urbana, IL 61801, USA}

\author[0000-0003-1659-7035]{Yue Shen}
\email{shenyue@illinois.edu}
\affiliation{Department of Astronomy, University of Illinois at Urbana-Champaign, Urbana, IL 61801, USA}
\affiliation{National Center for Supercomputing Applications, University of Illinois at Urbana-Champaign, Urbana, IL 61801, USA}

\author[0000-0001-8416-7059]{Hengxiao Guo} 
\affiliation{Shanghai Astronomical Observatory, Chinese Academy of Sciences, 80 Nandan Road, Shanghai 200030, People's Republic of China} 

\author[0000-0002-6404-9562]{Scott F.~Anderson}
\affiliation{Astronomy Department, University of Washington, Box 351580, Seattle, WA 98195, USA}

\author[0000-0002-0167-2453]{W.~N.~Brandt}
\affiliation{Department of Astronomy and Astrophysics, The Pennsylvania State University, University Park, PA 16802, USA}
\affiliation{Institute for Gravitation and the Cosmos, The Pennsylvania State University, University Park, PA 16802, USA}
\affiliation{Department of Physics, 104 Davey Laboratory, The Pennsylvania State University, University Park, PA 16802, USA}

\author[0000-0001-9920-6057]{Catherine~J.~Grier}
\affiliation{Department of Astronomy, University of Wisconsin-Madison, Madison, WI 53706, USA}  

\author[0000-0002-1763-5825]{Patrick B. Hall}
\affiliation{Department of Physics and Astronomy, York University, 4700 Keele St., Toronto, ON M3J 1P3, Canada}

\author[0000-0001-6947-5846]{Luis C.~Ho}
\affiliation{Kavli Institute for Astronomy and Astrophysics, Peking University, Beijing 100871, China}
\affiliation{Department of Astronomy, School of Physics, Peking University, Beijing 100871, China}

\author[0000-0002-0957-7151]{Yasaman Homayouni}
\affiliation{Department of Astronomy and Astrophysics, The Pennsylvania State University, University Park, PA 16802, USA}
\affiliation{Space Telescope Science Institute, 3700 San Martin Drive, Baltimore, MD 21218, USA}

\author[0000-0003-1728-0304]{Keith Horne}
\affiliation{SUPA Physics and Astronomy, University of St Andrews, Fife, KY16 9SS, UK}

\author[0000-0002-0311-2812]{Jennifer~I-Hsiu Li}
\affiliation{Michigan Institute for Data Science, University of Michigan, Ann Arbor, MI, 48109, USA}
\affiliation{Department of Astronomy, University of Michigan, Ann Arbor, MI, 48109, USA}

\author[0000-0001-7240-7449]{Donald P. Schneider}
\affiliation{Department of Astronomy and Astrophysics, The Pennsylvania State University, University Park, PA 16802, USA}
\affiliation{Institute for Gravitation and the Cosmos, The Pennsylvania State University, University Park, PA 16802, USA}


\begin{abstract}
Over three decades of reverberation mapping (RM) studies on local broad-line active galactic nuclei (AGNs) have measured reliable black-hole (BH) masses for $> 100$ AGNs. These RM measurements reveal a significant correlation between the Balmer broad-line region size and the AGN optical luminosity (the $R-L$ relation). Recent RM studies for AGN samples with more diverse BH accretion parameters (e.g., mass and Eddington ratio) reveal a substantial intrinsic dispersion around the average $R-L$ relation, suggesting variations in the overall spectral energy distribution shape as functions of accretion parameters. 
Here we perform a detailed photoionization investigation of expected broad-line properties as functions of accretion parameters, using the latest models for the AGN continuum implemented in {\tt qsosed}. We compare theoretical predictions with observations of a sample of 67 $z\lesssim0.5$ reverberation-mapped AGNs with both rest-frame optical and UV spectra in the moderate-accretion regime (Eddington ratio $\Redd\equiv L/L_{\rm Edd}<0.5$). The UV/optical line strengths and their dependences on accretion parameters can be reasonably well reproduced by the locally-optimally-emitting cloud (LOC) photoionization models. We provide quantitative recipes that use optical/UV line flux ratios to infer the ionizing continuum, which is not directly observable. In addition, photoionization models with universal values of ionization parameter ($\log U_{\rm H}=-2$) and hydrogen density ($\log n({\rm H})=12$) can qualitatively reproduce the observed global $R-L$ relation for the current AGN sample. However, such models fail to reproduce the observed trend of decreasing BLR size with $L/L_{\rm Edd}$ at fixed optical luminosity, which may imply that the gas density increases with the accretion rate.  


\end{abstract}


\section{Introduction}\label{sec:intro}


Reverberation mapping (RM) is a well-demonstrated technique to measure the black-hole (BH) masses in active galactic nuclei (AGNs), using the distance and kinematics of the broad line region (BLR) clouds. After several decades of RM work by different groups, there are now more than 100 local AGNs ($z\lesssim0.3$) that have successful mass measurements with \hbeta\ RM \citep[e.g.][]{Blandford&McKee_1982, Peterson1993, Peterson_etal_1998, Peterson_etal_2002, Peterson2004, Kaspi2007, Barth_etal_2015, U_etal_2022, Du2015_SEAMBH, Du2018_SEAMBH, Hu2021_SEAMBH, Malik_etal_2023, Woo_etal_2023}. Recent dedicated RM programs are also pushing these measurements to higher redshifts and to cover additional broad lines such as \CIV\ and \MgII\ \citep{Lira_etal_2018, Grier_etal_2019, Homayouni_etal_2020, Kaspi_etal_2021, Hoormann_etal_2019, Malik_etal_2023, Yu_etal_2021_OzDES}. Earlier studies on nearby RM AGNs have revealed a tight (i.e., scatter $\lesssim 0.15$~dex in BLR size) correlation between the average distance from the BH to \hbeta-emitting BLR clouds and the optical luminosity $\L5100$ of the AGN, known as the radius-luminosity ($R-L$) relation \citep{Kaspi2000, Kaspi2007, Bentz2013}. This $R-L$ relation provides the foundation for the so-called single-epoch (SE) virial BH mass recipes \citep{Vestergaard&Peterson2006, Shen2013} that use a single monochromatic continuum luminosity to estimate the time lag. Due to their simplicity, these single-epoch BH mass recipes have been widely adopted to estimate BH masses in high-redshift quasars and for large AGN samples with single-epoch spectra \citep[e.g.,][]{Shen_etal_2011, Wu&Shen2022}. Furthermore, this tight local $R-L$ relation motivates its potential utility as a standard candle to measure luminosity distances for cosmology \citep{Watson2011, Martinez-Aldama_etal_2019}, once the BLR size is measured directly via RM. 

However, recent RM studies targeting AGNs across broad ranges of luminosities and Eddington ratios have found evidence for increased dispersion around the canonical $R-L$ relation \citep[e.g.,][]{Du2014_SEAMBH, Grier_etal_2017, Fonseca_etal_2020, Shen_etal_2024}. In particular, the super-Eddington accreting massive black hole (SEAMBH) collaboration \citep{Du2014_SEAMBH, Du2015_SEAMBH, Du2018_SEAMBH, Hu2021_SEAMBH} as well as the Seoul National University (SNU) AGN monitoring project \citep{Woo_etal_2023} targeting high-accretion-rate AGNs have found a surprising downward offset of the average H$\beta$ lag from the canonical relation in these objects. The average time lags for these high-accretion-rate AGNs are consistently shorter than predictions from the canonical relation, and can reach as much as a factor of five times shorter in individual objects, which translates to a similar factor of offset in BH mass. The recent dynamical mass measurement of SDSS J092034.17+065718.0 ($z\approx 2.3$) from GRAVITY+ also reported a smaller BLR size than predicted by the canonical $R-L$ relation for this super-Eddington accretion quasar \citep{Abuter_etal_2024}. This spectro-interferometric method independently confirms that super-Eddington AGNs may have smaller BLR sizes as measured from reverberation mapping. Meanwhile, the Sloan Digital Sky Survey Reverberation Mapping (SDSS-RM) project \citep{Shen_etal_2015, Shen_etal_2024, Blanton_etal_2017} targeting non-local AGNs at $0.1<z<4.5$ also found a large dispersion of $\sim 0.3$ dex around the mean \hbeta\ $R-L$ relation.

This increased scatter in the $R-L$ relation has gained significant interest in recent work \citep{Czerny_etal_2019, Du&Wang2019, Dalla_etal_2020, Fonseca_etal_2020}. Understanding the origin of this dispersion and deriving corrections to potentially ``tighten'' the $R-L$ relation would have tremendous value both in designing better single-epoch mass recipes and in using the $R-L$ relation as a luminosity indicator. Detailed cadence and duration simulations of RM monitoring have ruled out selection bias as a major contributor to the observed scatter \citep{Li_etal_2009, Fonseca_etal_2020}. Several recent works have attempted to use the optical \FeII strength (denoted by $R_{\rm FeII} \equiv {\rm EW}_{\rm FeII,4434-4684}/{\rm EW}_{\rm H\beta}$) as a proxy to correct this offset \citep{Du&Wang2019, Martínez-Aldama_etal_2020}, where $R_{\rm FeII}$ is considered a proxy of the Eddington ratio through the Eigenvector 1 relations \citep{Boroson&Green1992, Shen&Ho2014}. 

In this work, we consider the standard scenario that the BLR emission is sensitive to the ionizing continuum, which depends on the accretion rate and consequently, the structure of the accretion flow. 
Accretion flows around black holes can be classified into three main regimes based on their accretion rates (here we use the dimensionless Eddington ratio $\Redd \equiv L_{\rm bol}/L_{\rm Edd} = L_{\rm bol}/(1.26 \times 10^{38} \Mbh/M_\odot)$).
Firstly, at low-to-moderate accretion rates ($\Redd \lesssim 0.5$), the accretion flow can be reasonably modeled by the geometrically thin Shakura-Sunyaev Disk \citep[SSD,][]{Shakura&Sunyaev1973}, where the optically-thick gas radiates multi-temperature blackbody radiation.
As the accretion rate substantially increases to $\Redd\gtrsim1$, the structure of the accretion flow undergoes significant transformations due to increased radiation pressure, leading to a slim disk \citep{Abramowicz1988} where the disk height becomes comparable to the disk radius. Many of the high-Eddington-ratio AGNs included in the SEAMBH sample likely are better described by this slim disk model \citep{Du2014_SEAMBH}.
Conversely, at very low accretion rates ($\Redd < 10^{-3}$), the viscously dissipated energy heats the flow rather than being radiated away, resulting in an optically-thin, hot advection-dominated accretion flow \citep[ADAF,][] {Yuan&Narayan2014}. 
Recognizing that the black-hole accretion rate in different regimes can lead to dramatic changes in the accretion-flow geometry and the resulting shape of the spectral energy distribution (SED) \citep[e.g.,][]{Ho2008, Castello_etal_2016}, and given that the exact transition \Redd\ between different disk geometry remains uncertain, we focus on AGNs with low-to-moderate accretion rates ($10^{-3}<\Redd<0.5$), where the accretion flow is predominately in the SSD regime. However, contributions from the hot corona (and a potential warm corona) are also included in the model continuum SEDs (see \S\ref{sec:SED} for details). 


To fully understand the connection between the underlying ionizing continuum and observables, detailed photoionization calculations are required. In this work, we construct a series of SEDs and photoionization calculations to compare with the average \hbeta\ emission distance from RM and the observed quasar spectral properties. In \S\ref{sec:theory}, we describe our theoretical models for the AGN SEDs and our photoionization calculations using the LOC model \citep{Baldwin_etal_1995}. In \S\ref{sec:obs}, we present our analysis of the observational data and comparison with the theoretical framework. We discuss our results in \S\ref{sec:discussion} and conclude in \S\ref{sec:con}. Throughout this paper we adopt a flat $\Lambda$CDM cosmology with $\Omega_M=0.3$ and $H_0=70\,\rm{km\,s^{-1}Mpc^{-1}}$.

\section{Theoretical framework}\label{sec:theory}

A physically motivated photoionization model for the AGN BLR is the locally optimally emitting cloud (LOC) model \citep{Baldwin_etal_1995}. This model assumes an axisymmetric distribution of continuous clouds with varying gas densities and distances from the central continuum source. The observed line emission is the collection of all illuminated clouds but is dominated by those with the highest efficiency of reprocessing the incident ionizing continuum. In the following section, we apply the LOC model to a series of SEDs with different BH parameters (mass and accretion rate), to predict broad-emission-line properties with photoionization calculations. 

\subsection{AGN SEDs with XSPEC}\label{sec:SED}

\begin{figure}
\centering
    \includegraphics[width=\linewidth]{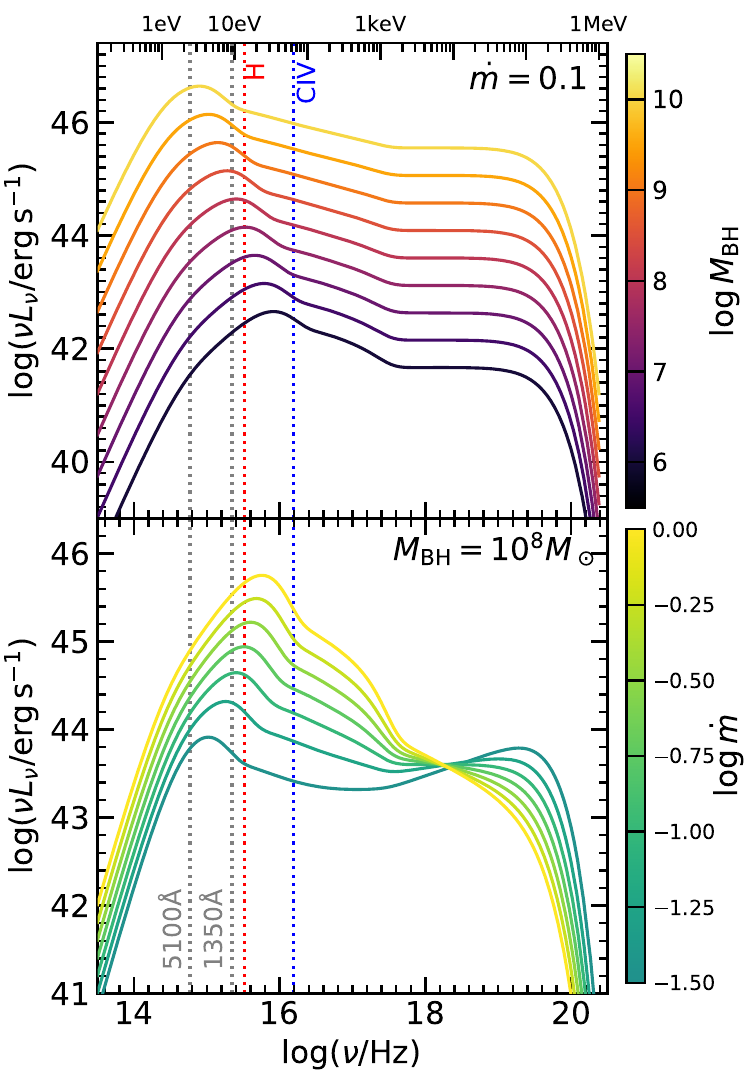}
    \caption{Example of output SEDs generated using {\tt Xspec qsosed}. The vertical blue and red dotted lines represent photon energies of $h\nu_{\rm CIV}=64.5\,{\rm eV}$ and $h\nu_H=13.6\,{\rm eV}$, respectively. Top Panel: SEDs with a constant accretion rate ($\dotm=0.1$) and varying BH mass ($\Mbh=10^{6-10}M_\odot$) in logarithmic intervals of 0.5 dex. Bottom Panel: SEDs with fixed BH mass ($M_{\rm BH}=10^{8}M_\odot$) while varying the accretion rate $\log\dot{m}$ from $-1.5$ to $0$ in logarithmic intervals of 0.25 dex.}
    \label{fig:SED}
\end{figure}

\begin{figure*}
    \includegraphics[width=\linewidth]{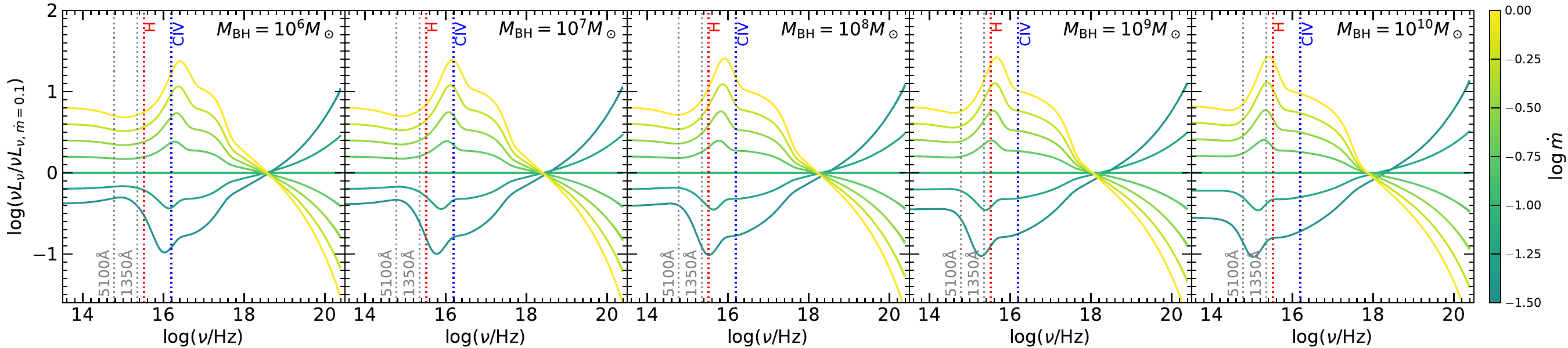}
    \caption{SEDs computed for various accretion rates $\log\dot{m}\in[-1.5, 0]$ with 0.25 dex intervals at constant BH mass \Mbh\ (from left to right: $10^6$, $10^7$, $10^8$, $10^9$, and $10^{10}\,M_\odot$). Each SED within the panel is normalized relative to the SED with \dotm=0.1. Vertical lines are placed at photon energies of $h\nu_{\rm CIV}=64.5$ eV (blue) and $h\nu_H=13.6$ eV (red), respectively. }
    \label{fig:SED_mass}
\end{figure*}

\begin{figure}
    \includegraphics[width=\linewidth]{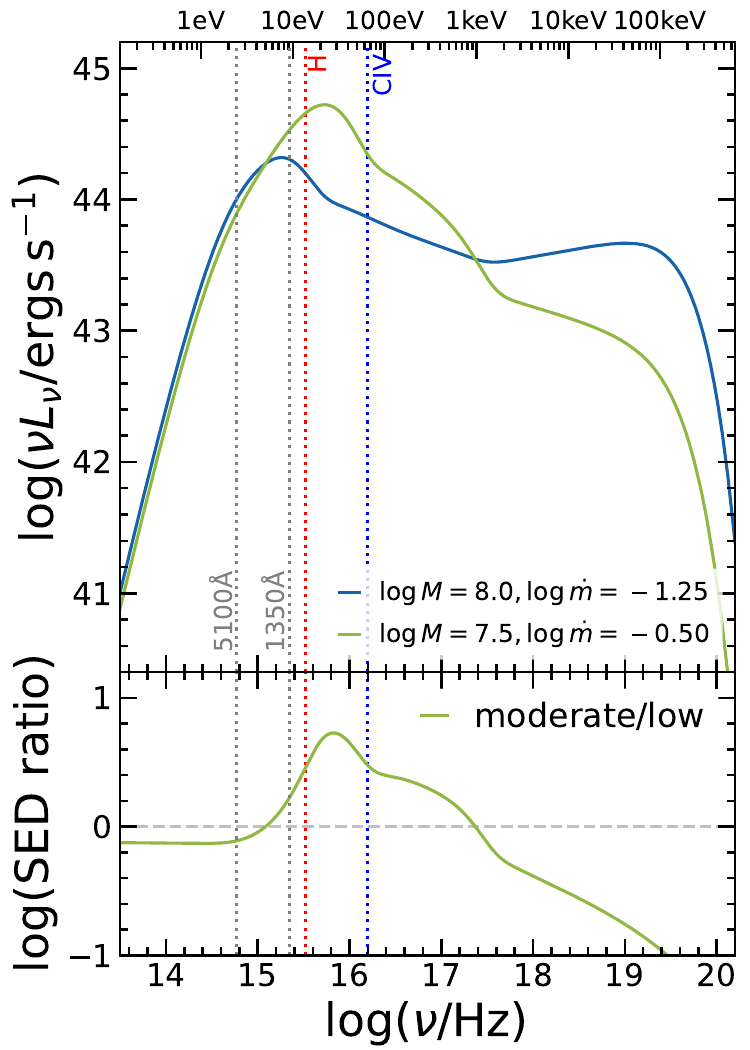}
    \caption{SEDs of two SMBHs with mass \Mbh\ of $10^{7.5}$ and $10^8 M_\odot$ and accretion rate $\log\dot{m}$ of $-0.5$ and $-1.25$, respectively. These two SEDs with diverse SED shapes produce similar optical luminosities $\log\L5100\sim 44$ but different ionizing fluxes. Within the SSD regime and at fixed optical luminosity, AGNs with higher Eddington ratios (lower BH masses) have relatively more ionizing flux compared with the 5100\,\AA\ flux, which would presumably lead to longer BLR lags.  
    }
    \label{fig:SED_fix5100}
\end{figure}

\begin{figure}
    \includegraphics[width=\linewidth]{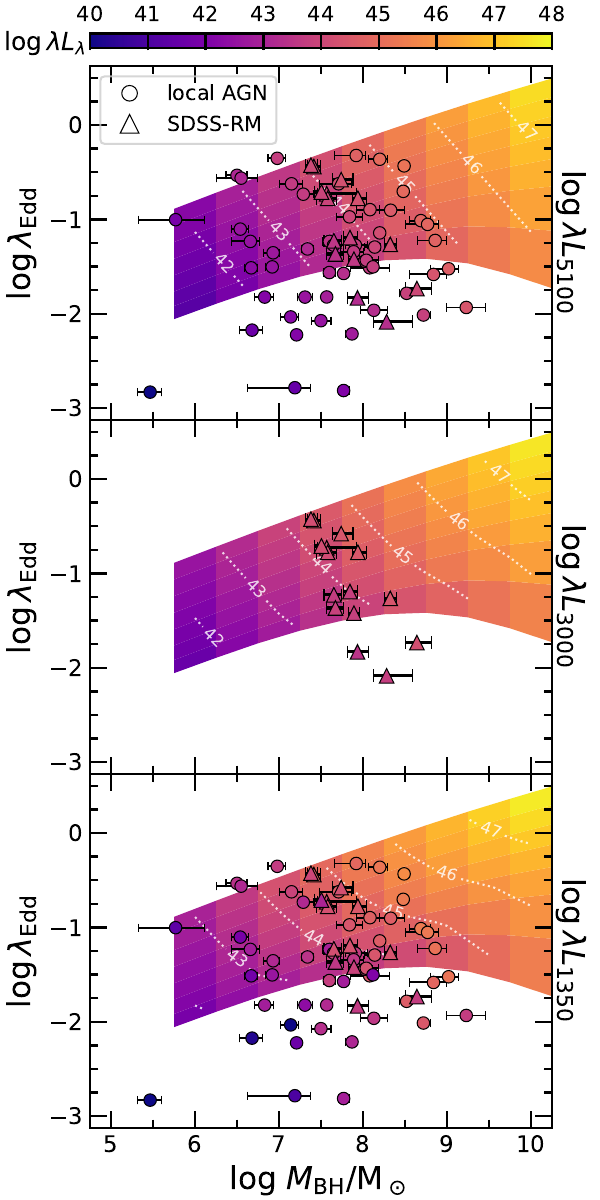}
    \caption{Continuum luminosities at $5100{\rm \AA}$ (top), $3000{\rm \AA}$ (middle) and $1350{\rm \AA}$ (bottom) in the \Mbh-\Redd\ plane, computed from the {\tt qsosed} models. Observed continuum luminosities of the compiled RM AGN sample are overlaid in each corresponding panel for comparison. Note that $\Redd<-1.5$ for certain theoretical SEDs is due to the constant bolometric correction applied to the continuum luminosity at 5100\,\AA. 
    }\label{fig:Lconti_grid}
\end{figure}


To generate a grid of ionizing SEDs with various BH parameters, we use the {\tt qsosed} \citep{Kubota&Done2018} model in {\tt Xspec} \citep{Xspec} that has three featured components: one truncated standard thin-disk component providing multi-temperature blackbody radiation; a hot Comptonized corona for non-thermal power-law emission dominating the X-ray continuum above 1 keV; and a soft X-ray excess from an intermediate warm Comptonizing component. 
Compared to its predecessor {\tt OPTXAGNF} \citep{Done_etal_2012}, which assumes a single Comptonization spectrum for the seed photon energy due to the lack of interstellar absorption observations, the updated, physically driven model {\tt qsosed} treats the warm Comptonization component more carefully and assumes that the seed photon energy originates from the underlying cool material in the mid-plane \citep{Kubota&Done2018}. Both \cite{Panda_etal_2019} and \cite{Ferland_etal_2020} utilize the {\tt qsosed} model in their BLR photoionization computations to explore the impact of the Eddington ratio on the BLR photoionization and demonstrate consistent results with observations. Moreover, comparison with the stacked AGN spectra shows that the {\tt qsosed}-predicted SEDs align well with intermediate black hole masses in the range of $7.5<\log\Mbh<9.0$ while showing discrepancy with observations at optical/EUV for both ends \citep{Mitchell_etal2023}. 
The uncertainties and limitations of {\tt qsosed} will be further discussed in \S\ref{sec:uncer}.
The {\tt qsosed} model fixes several disk parameters and includes reprocessing; the free input parameters are: the BH mass \Mbh, the co-moving distance from the object to the observer ($D_c$), the Eddington-scaled accretion rate (\dotm$ \equiv \dot{M}/\dot{M}_{\rm Edd} = \eta\dot{M}c^2/L_{\rm Edd}$, where $\eta$ is the spin-dependent radiative efficiency, fixed at 0.057 for a non-spinning black hole in {\tt qsosed}), the dimensionless spin parameter ($a_*$), the cosine of the inclination angle for the warm Comptonising component and the outer disk ($\cos{i}$), and the redshift ($z$). As an exploratory study, we consider non-spinning ($a_*=0$) BHs with a mass range of $10^{6}-10^{10} \rm{M_\odot}$ and accretion rates $-1.5\leq\log\dot{m}\leq0$, with $9\times7$ logarithmically-space grid points in each direction and intervals of $0.5$ and $0.25$ dex. We fix the co-moving distance $D_c=1000\,{\rm Mpc}$, redshift $z=0$, and the inclination at $\cos{i}=0.5$ to compute the source SEDs on the \Mbh-\dotm\ plane.

We focus exclusively on the moderate accretion rate regime of $\log\dotm\in[-1.5, 0] $, because of substantial theoretical uncertainties inherent in the slim disk SED model, as well as the imposed lowest accretion rate limit of $-1.65<\log\dotm$ in {\tt qsosed}.
To compare with observables, we do not use the mass-accretion rate (\dotm) for these theoretical SEDs but instead use the Eddington ratio (\Redd) based on the bolometric luminosity calculated using the $5100 {\rm \AA}$ monochromatic luminosity. This detail would lead to \Redd\ values at $\log \lambda_{\rm Edd}<-1.5$ for our theoretical SEDs. 




Several example output SEDs from {\tt qsosed} are displayed in Figure~\ref{fig:SED}. The optical and UV continuum wavelengths used for fiducial luminosity calculations are marked by the grey dashed vertical lines, and the ionizing energy for \CIV\ and hydrogen are shown in blue and red vertical dotted lines, respectively ($h\nu_{\rm CIV}=64.5{\rm eV}$ and $h\nu_H=13.6{\rm eV}$). The top panel illustrates how the SED varies as the BH mass changes at a fixed accretion rate $\dotm=0.1$. The lower panel displays the SEDs variation as a function of accretion rate at a fixed BH mass $\Mbh=10^8M_\odot$. Figure~\ref{fig:SED} illustrates that the increases in both BH mass and accretion rate result in increased optical and UV continuum luminosities ($\L5100$ and $\lambda L_{1350}$), accompanied by an increase in the ionizing luminosity. The ratio of line-ionizing fluxes \CIV/hydrogen changes with the BH mass and accretion rate as well.

Figure~\ref{fig:SED_mass} shows SED comparisons for accretion rates $\dotm\in[-1.5, 0]$ at fixed mass ($10^6$, $10^7$, $10^8$, $10^9$ and $10^{10}\, M_\odot$). At each BH mass, the SEDs are normalized by the SED with $\log\dotm=-1$. In all panels, higher accretion rates have relatively higher optical and UV luminosities and overall more hydrogen ionizing flux. Meanwhile, the change in UV luminosity $\lambda L_{1350}$ is more sensitive to accretion rate than $\L5100$, especially for more massive black holes. To illustrate the significant variations in the SED due to accretion rate changes, Figure~\ref{fig:SED_fix5100} presents two theoretical SEDs with similar optical luminosities $\log\L5100\sim 44$ but different accretion parameters (mass and accretion rate). Although the optical luminosity is similar, the more massive BH with a lower \Redd\ (in blue) produces significantly less hydrogen ionizing flux and hence is expected to have a shorter \hbeta\ time lag, if assuming the localization of the BLR is related to the ionizing flux and other properties of the BLR are the same for the two cases. 

Figure~\ref{fig:Lconti_grid} displays the continuum luminosities computed from {\tt qsosed} SEDs at three optical/UV wavelengths, with slightly higher luminosities at shorter wavelengths. While the simulated mass-accretion rate ($\log \dot{m}$) ranges from $-1.5$ to 0 in {\tt qsosed}, the \Redd\ values may fall below $-1.5$ for the theoretical SEDs due to the universal bolometric correction (${\rm BC}=9.26$) applied to the continuum luminosity at 5100\,\AA. 
The continuum luminosities show a diagonal increasing trend on the \Mbh-\Redd\ plane. This trend suggests that despite variations in their underlying SED shapes, BHs with different accretion parameters can produce similar continuum luminosities at fixed observed wavelengths.

This exploratory investigation demonstrates the possible diverse SEDs from AGNs with different accretion parameters. Since the BLR gas clouds are ionized by the ionizing continuum flux, the qualitative changes in the SED shape conceivably can lead to dispersion in the observed $R-L$ relation based on the optical luminosity. We now proceed to use photoionization calculations to make quantitative predictions. 

\subsection{Photoionization with CLOUDY} \label{sec:photoionization}


We construct detailed photoionization models to calculate line luminosities for individual clouds using {\tt CLOUDY} (\cite{CLOUDY}, version 17.02) with a spherically symmetric geometry to account for various excitation mechanisms and radiative transfer effects. SEDs computed from \S\ref{sec:SED} are used as the incident radiation field input. This approach assumes the BLR gas sees all the radiation from the central source, i.e., there is no inner shielding of the radiation. 

Following previous works in a spherically symmetric geometry \citep{Korista&Goad_2000, Korista&Goad_2004, Guo_etal_2020}, we assume that the BLR clouds cover a broad range of hydrogen density ($6.0\leqslant\log{n(\rm{H})}\leqslant14.0$) and surface ionization flux ($16.0\leqslant\log{\Phi(\rm{H})}\leqslant25.0$), both with logarithmic intervals of 0.125 dex. The ionization parameter $U_{\rm H}$ is defined as a dimensionless ratio between the hydrogen-ionizing photon flux $Q(\rm{H})$ and the total hydrogen density $n(\rm{H})$:
\begin{equation}\label{eq:ionizing_para}
    U_{\rm H} \equiv \frac{Q({\rm H})}{4\pi R_{\rm BLR}^2 n({\rm H})c}\equiv\frac{\Phi(\rm{H})}{n({\rm H})c},
\end{equation}
where $R_{\rm BLR}$ is the distance between the central ionizing source and the illuminated surface of the cloud and $c$ is the speed of light. For each cloud, we assume a hydrogen column density $N_{\rm H} = 10^{23}\, {\rm cm ^{-2}}$, abundance $Z=Z_\odot$, and an overall covering factor of $CF=50\%$. Given the full set of hydrogen densities and ionizing flux, we compute the photoionization results with 4745 calculations for each SED.

The total line luminosity is computed using the LOC model by summing over all grid points with proper weights determined from assumed distribution functions. When summing the grid emissivity, only ionized clouds with a density range of $8.0\leqslant\log{n(\rm{H})}\leqslant12.0$ and ionizing flux $6.0\leqslant\log{U_{\rm H} c}\leqslant11.25$ are considered to obtain the radial surface emissivity ($F(r)$) \citep[see details in][]{Korista&Goad_2000, Korista&Goad_2004, Korista&Goad2019, Guo_etal_2020}. For simplicity, we adopt the empirical assumptions outlined in \cite{Baldwin_etal_1995}: $f(r)\propto r^{\Gamma}$ and $g(n)\propto n^{\beta}$ ($\beta=-1$) for the cloud distribution, which represent the cloud coverage fractions as functions of radius and gas density, respectively. This assumption ensures equal weighting for each grid point in the density-flux plane on a logarithmic scale. In previous studies, \cite{Korista&Goad_2000} and \cite{Korista&Goad2019} proposed $-1.4<\Gamma<-1$ and fix $\Gamma=-1.2$ for NGC 5548; \cite{Guo_etal_2020} employed $\Gamma =-1.1$ to align with the observed \MgII\ luminosity; therefore, after comparing with observed line luminosities in \S\ref{sec:obs_line_strength}, we fix $\Gamma=-1.1$ to match the observed lines. The observed line luminosity is:
\begin{equation}\label{eq:LOC_integral}
\begin{split}
    L_{\rm line} & \propto\int\int^{R_{\rm out}}_{R_{\rm in}}r^{2}F(r)f(r)g(n) dn dr,  \\
     & \propto \int\,d(\log{n})\int^{R_{\rm out}}_{R_{\rm in}} r^{1.9}F(r)\,d(\log{r}),
\end{split}
\end{equation}
where $F(r)$ is the radial surface emissivity of a single cloud. The BLR sizes predicted in \S\ref{sec:predict_R-L} are adopted as the average distance for the integral.

\subsection{Predictions of BLR size}\label{sec:predict_R-L}


\begin{figure*}
    \includegraphics[width=\linewidth]{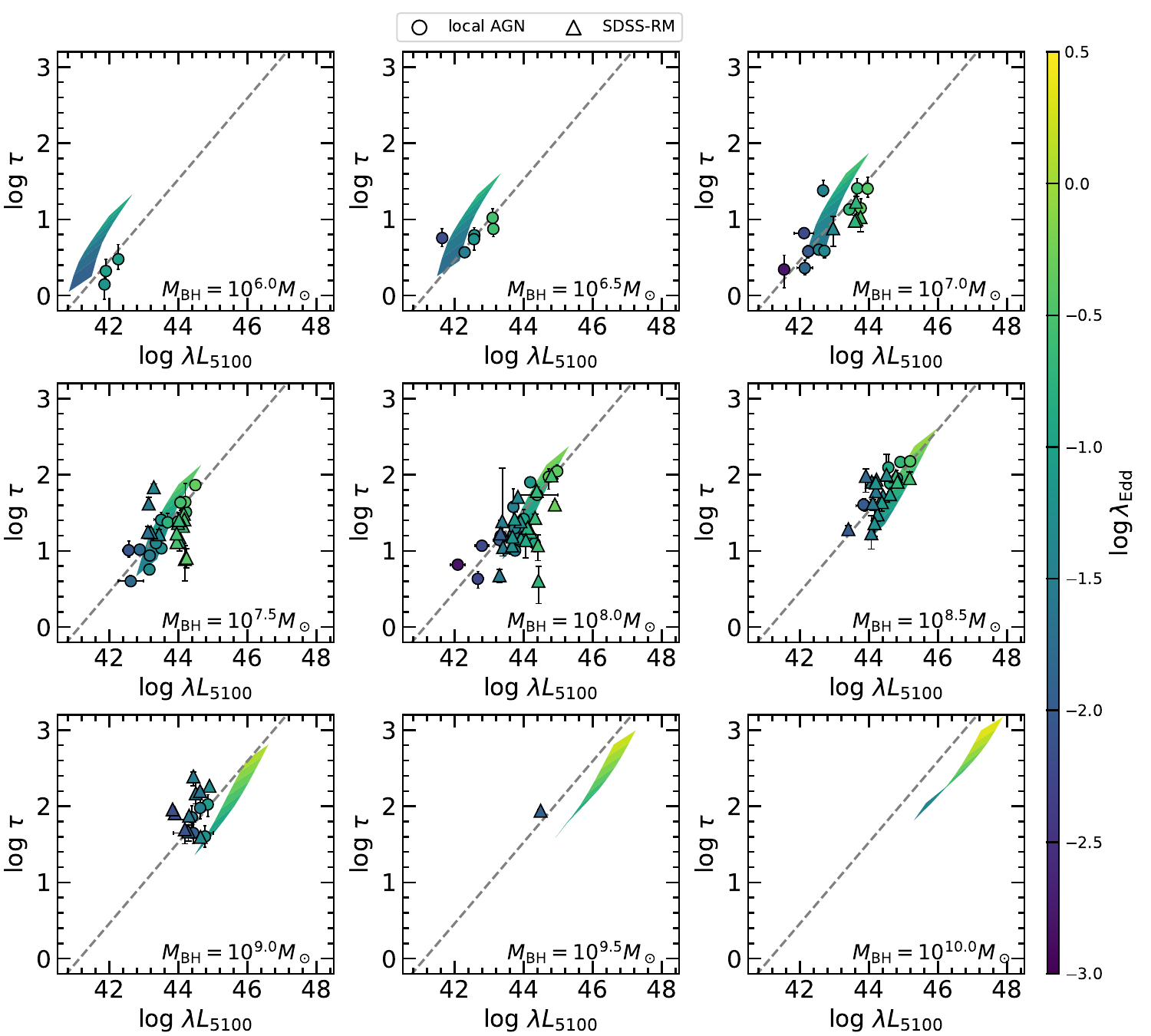}
    \caption{Dependence of the \hbeta\ BLR size on monochromatic luminosity at 5100 ${\rm \AA}$ across a range of mass bins (from $\log\Mbh=6$ to $\log\Mbh=10$ with $\Delta\log\Mbh=0.5$). Predicted time lags are represented by shaded bands with varying filling colors denoting different Eddington ratios. Observed \hbeta\ time lags are presented as circles and triangles for low accretion ($\Redd<0.5$) local AGNs and SDSS-RM sources, respectively, with the same color scheme as the theoretical predictions. The grey dashed line indicates the canonical $R-L$ relation from \cite{Bentz2013}, with a slope $k = 0.533$ and an intercept $b=1.527$ at $\log\lambda L_{5100}=44$.}
    \label{fig:theory_R_L_grid}
\end{figure*}

The relation between BLR radius $R_{\rm BLR}=c\tau$ and monochromatic luminosity can be calculated using this grid of SEDs. We use the following expression to compute the average distance between the ionizing source center and the Balmer-line-emitting clouds:
\begin{equation}\label{eq:R_BLR}
    R_{\rm BLR} = \sqrt{\frac{Q({\rm H})}{4\pi\Phi({\rm H})}} = \sqrt{\frac{\int^\infty_\nu \frac{\pi L_\nu}{h\nu}d\nu}{4\pi\Phi({\rm H})}}
\end{equation}
where $\Phi({\rm H})=U_{\rm H}n({\rm H})c$ is the flux of ionizing photons received by the clouds. 
The LOC model assumes that the clouds with the highest reprocessing efficiency dominate the observed emissions. Hydrogen and Helium emission lines, in particular, exhibit peak reprocessing efficiency under conditions of low ionizing photon flux and high hydrogen density ($\log n({\rm H})\gtrsim12$) \citep[e.g.,][]{Baldwin_etal_1995, Korista_etal_1997, Korista&Goad_2000}. Moreover, comparisons of several AGN emission line ratios with the photoionization of the dominant cloud have consistently indicated a product of the ionization parameter and hydrogen density approximating $\log U_{\rm H} n({\rm H}) \sim 10$ ($\log n({\rm H}) \sim 11.6$ to $12.6$ and $\log U_{\rm H}\sim-2.5$ to $-2.0$) \citep[e.g.,][]{Signut&Pradhan2003, Matsuoka_etal_2008, Negrete_etal_2012}. Therefore, constant ionization parameter $\log U_{\rm H}=-2.0$ and hydrogen density $n(\rm{H})= 10^{12}\,{\rm cm^{-3}}$ are adopted here to calculate the BLR gas distance (Eqn.~\ref{eq:R_BLR}).

Figure~\ref{fig:theory_R_L_grid} compares our predicted \hbeta\ BLR size as a function of the monochromatic luminosity at $5100{\rm \AA}$ with our compiled RM AGN sample (see details in \S\ref{sec:obs}). We divide the observed RM sample into narrow mass bins with $\Delta\log\Mbh=0.5$ and assign them to their respective mass bins in Figure~\ref{fig:theory_R_L_grid}. Within each subplot, the predicted time lags are depicted by colored bands, and all data points are color-coded based on their Eddington ratio. As shown in Figure~\ref{fig:theory_R_L_grid}, our predicted time lag trends with optical luminosity are in overall agreement with the observed lags given the assumed universal values of the ionization parameter and cloud hydrogen density. This consistency supports the general notion that the BLR responds to the ionizing continuum. While the assumed photoionization parameters could reproduce the overall trend in the $R-L$ plane across a range of BH mass, discrepancies in measured accretion parameters and/or assumptions in our theoretical calculations for individual objects may lead to deviations from the predicted time lags; these observational and theoretical uncertainties are discussed in \S\ref{sec:discussion}.

\section{Comparison with observations}\label{sec:obs}

\subsection{Sample and data}

\begin{figure}
\centering
    \includegraphics[width=\linewidth]{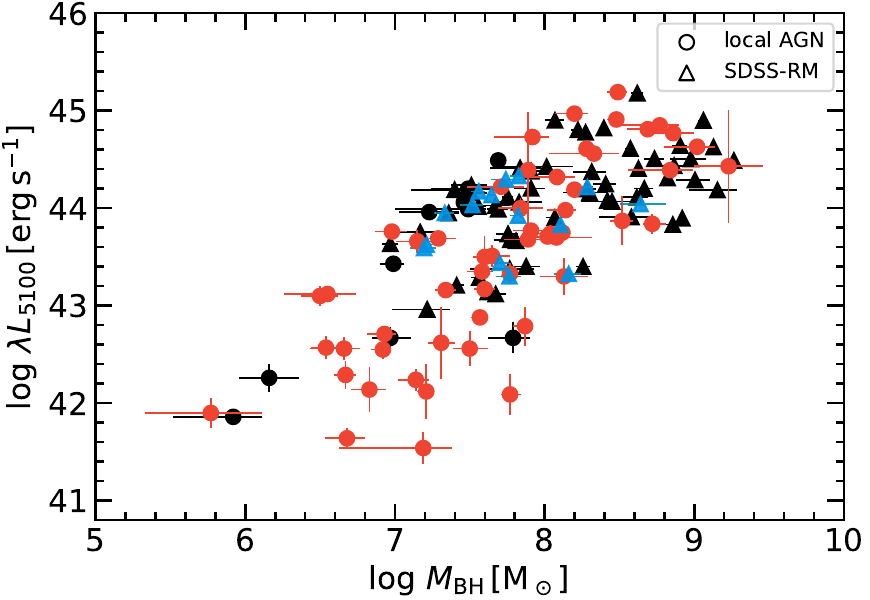}
    \caption{Black hole mass measurements from \hbeta\ RM and optical luminosity $\L5100$ for local AGNs at $z<0.5$ are represented by circles \citep{Du&Wang2019}, while the SDSS-RM sample is coded by triangles \citep{Grier_etal_2017, Shen_etal_2019b, Shen_etal_2024}. Objects lacking ultraviolet spectra are indicated in black, whereas those with ultraviolet spectra are distinguished by colors, with local AGNs in red and the SDSS-RM sample in blue.
    }\label{fig:MBHvsLOGL5100}
\end{figure}

\begin{figure*}
    \includegraphics[width=\linewidth]{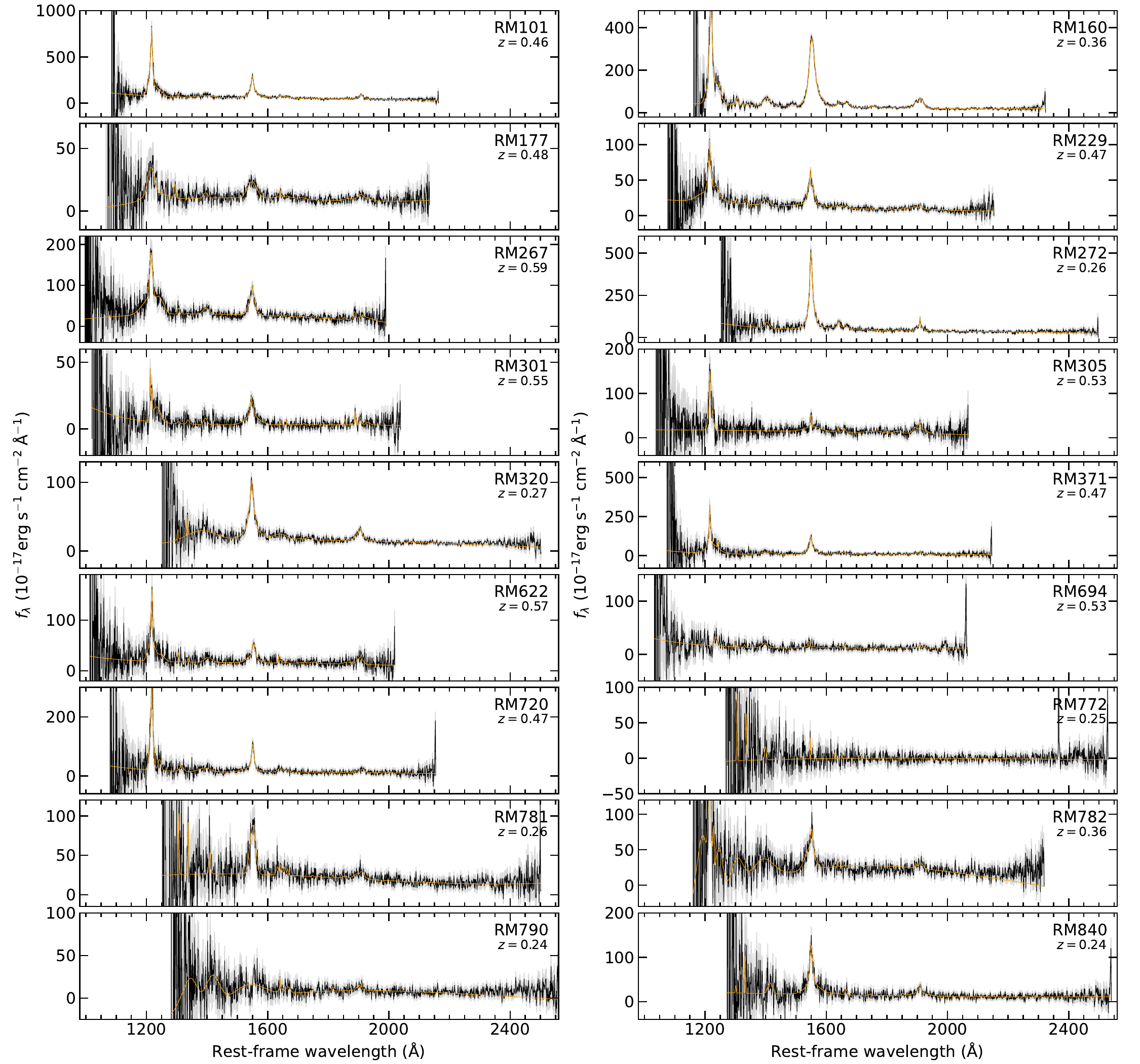}
    \caption{UV spectra of the 18 quasars observed by HST/STIS under the program HST GO-16171. The data have been corrected for the galactic reddening and shifted to the rest frame. The orange lines demonstrate our fitting results.}
    \label{fig:hst_stis_spec}
\end{figure*}

\begin{table*}
\caption{Target Properties}
\label{tab:SDSSRM-sample}
\begin{tabular}{ccccccc}
\hline\hline
RMID & R.\,A. (J2000) & Dec. (J2000) & $z$ & $\tau_{\rm H\beta}$ & log(${{M}_{\rm BH, RM}}$) & NUV \\
 & (deg) & (deg) &   & (days) & ($\rm M_{\odot}$) & (mag) \\
\hline\hline
RM101 & 213.059 & 53.430 & 0.458 & $21.4^{+4.2}_{-6.4}$ & $7.26^{+0.09}_{-0.12}$ & 19.35  \\ 
RM160 & 212.672 & 53.314 & 0.360 & $21.9^{+4.2}_{-2.4}$ & $7.85^{+0.09}_{-0.07}$ & 20.77  \\ 
RM177 & 214.352 & 52.507 & 0.482 & $10.1^{+12.5}_{-2.7}$ & $7.57^{+0.35}_{-0.11}$ & 20.80 \\ 
RM229 & 212.575 & 53.494 & 0.470 & $16.2^{+2.9}_{-4.5}$ & $7.65^{+0.09}_{-0.12}$ & 20.77 \\ 
RM267 & 212.803 & 53.752 & 0.588 & $20.4^{+2.5}_{-2.0}$ & $7.41^{+0.08}_{-0.08}$ & 20.16 \\ 
RM272 & 214.107 & 53.911 & 0.263 & $15.1^{+3.2}_{-4.6}$ & $7.58^{+0.10}_{-0.13}$ & 19.40 \\ 
RM301 & 215.043 & 52.675 & 0.548 & $12.8^{+5.7}_{-4.5}$ & $8.64^{+0.17}_{-0.14}$ & 21.10 \\ 
RM305 & 212.518 & 52.528 & 0.527 & $53.5^{+4.2}_{-4.0}$ & $8.32^{+0.07}_{-0.07}$ & 20.28 \\ 
RM320 & 215.161 & 53.405 & 0.265 & $25.2^{+4.7}_{-5.7}$ & $7.67^{+0.09}_{-0.11}$ & 20.60 \\ 
RM371 & 212.848 & 52.225 & 0.473 & $13.0^{+1.4}_{-0.8}$ & $7.38^{+0.08}_{-0.07}$ & 20.21 \\ 
RM622 & 212.813 & 51.869 & 0.572 & $49.1^{+11.1}_{-2.0}$ & $7.94^{+0.11}_{-0.06}$ & 20.55 \\ 
RM694 & 214.278 & 51.728 & 0.532 & $10.4^{+6.3}_{-3.0}$ & $6.70^{+0.20}_{-0.15}$ & 20.83 \\ 
RM720 & 211.325 & 53.258 & 0.467 & $41.6^{+14.8}_{-8.3}$ & $7.74^{+0.14}_{-0.10}$ & 19.80 \\ 
RM772 & 215.400 & 52.527 & 0.249 & $3.9^{+0.9}_{-0.9}$ & $6.60^{+0.10}_{-0.10}$ & 19.92 \\ 
RM781 & 215.265 & 51.972 & 0.264 & $75.2^{+3.2}_{-3.3}$ & $7.89^{+0.07}_{-0.07}$ & 20.47 \\ 
RM782 & 213.329 & 54.534 & 0.363 & $20.0^{+1.1}_{-3.0}$ & $7.51^{+0.06}_{-0.09}$ & 19.50 \\ 
RM790 & 214.372 & 53.307 & 0.238 & $5.5^{+5.7}_{-2.1}$ & $8.28^{+0.31}_{-0.15}$ & 20.10 \\ 
RM840 & 214.188 & 54.428 & 0.244 & $5.0^{+1.5}_{-1.4}$ & $7.93^{+0.13}_{-0.12}$ & 19.94 \\ 
\hline
\end{tabular}
\tablecomments{Properties of the 18 quasars proposed in HST GO-16171. RM black hole masses are based on \hbeta\ lags from \cite{Grier_etal_2017}. Time lags are rest-frame values.}
\end{table*}


We start from the sample of {$\sim 140$} $z\lesssim0.5$ broad-line AGNs with \hbeta\ RM measurements, and select those with $10^{-3}<\Redd<0.5$ in the SSD regime as our primary sample. Figure~\ref{fig:MBHvsLOGL5100} shows the distribution of BH mass \Mbh\ measured from \hbeta\ RM versus their optical luminosity $\L5100$ for this primary sample, including objects with and without UV spectroscopy. For \Mbh\ values of these local RM AGNs, we combine the catalog in \cite{Bentz&Katz_2015} and Table 1 in \cite{Du&Wang2019}, and correct \Mbh\ using a nominal virial factor $f_{\rm FWHM, H\beta}=1.12$ from \cite{Woo_etal_2015}. The host contamination in $\L5100$ has been subtracted for the local AGNs, except for eight sources (Zw~229+015, Mrk~1501, H0507+164, Mrk~704, PG 0934+013, Mrk~50, NGC~4395 and Mrk~6), for which we adopt their average $\L5100$ from light curves as an upper limit. We supplement the local RM AGNs with 72 more distant quasars from the SDSS-RM sample \citep{Shen_etal_2024}. The host contamination for the SDSS-RM quasars used in this work is removed based on a spectral decomposition approach described in \cite{Shen_etal_2015b}. We use a bolometric correction of ${\rm BC}=9.26$ \citep{Richards_etal_2006} for the optical continuum at $5100$~\AA\ to calculate the bolometric luminosity $L_{\rm bol}$, and the RM mass \Mbh\ for the dimensionless Eddington ratio. 

After assembling the primary sample of 135 low-to-moderate accretion AGNs with \hbeta\ RM measurements, we select a subsample of quasars with both rest-frame UV and optical spectra. We acquired UV spectroscopy for 18 SDSS-RM quasars at $0.2 \lesssim z \lesssim 0.6$ from a dedicated HST \textit{Space Telescope Imaging Spectrograph} (STIS) program in Cycle 28 (GO-16171; PI: Shen). These quasars are selected from the SDSS-RM sample \citep{Shen_etal_2015} with direct RM lags and BH masses based on the broad \hbeta\ line \citep{Grier_etal_2017}. We restrict to quasars with $0.2\lesssim z\lesssim 0.6$ and GALEX NUV $\lesssim 21$ mag while spanning a broad range in luminosity and \hbeta\ lags to sample the parameter space. Table \ref{tab:SDSSRM-sample} summarizes the properties of the 18 SDSS-RM targets.

The eighteen SDSS-RM quasars were observed with HST/STIS in spectroscopic mode with the NUV-MAMA G230L grating to optimize the spectral coverage. This instrumental configuration provides broad spectral coverage from 1570–3180 ${\rm\AA}$, which covers the dominant UV emission lines from \SiIV\ to \CIII\ for our targets. 
The brightnesses of our targets vary with GALEX NUV$\sim 19-21$: for the brightest twelve targets, one full orbit exposure time is utilized; five quasars (RM160, RM177, RM229, RM320, and RM694) were observed for two orbits; the faintest target with GALEX NUV magnitude of 21.1 (RM301) was observed with three orbits.
The initial visits of all targets lasted from Dec 2020 to Feb 2022; however, four targets (RM177, RM229, RM320, and RM694) suffered from the failure of guide-star acquisition at the start of the second orbit, resulting in lost data. Repeat visits for those targets were approved and recurred from September to November in 2022. The pipeline-processed STIS spectra of those targets with multiple orbits of exposure were coadded to improve the S/N of our sample. The observed spectra are presented in Figure~\ref{fig:hst_stis_spec}.

For the remaining local RM AGNs in our sample, in addition to our SDSS-RM quasars, we collect archival UV spectroscopy from the \textit{International Ultraviolet Explorer} (IUE), the STIS, and the \textit{Cosmic Origins Spectrograph} (COS) on HST. If multiple spectra exist in the public archives for a given object, either multiple epochs taken with the same instrument or from multiple instruments, we combined the spectra for each object via the standard weighted average method. Since the disk structure of high-accretion SMBHs will have a dramatic impact on the ionizing continuum illuminating the BLR, we only focus on the low-to-moderate accretion sample in this work. 

Our final sample includes 15 SDSS-RM quasars and 52 local RM AGNs with accretion rates $10^{-3}<\Redd<0.5$, 
We then apply the spectral decomposition code {\tt PyQSOFit} \citep{PyQSOFit} with minor custom adjustments to measure their UV spectral properties and summarize them in Table~\ref{tab:UV_spec_prop}. The optical spectral properties ($\log\L5100$, \hbeta\ and \HeIIopt) for the SDSS-RM sample are from \cite{Shen_etal_2019b}, while those of local AGNs are from \cite{Du&Wang2019}. 

\subsection{UV/optical line strengths}\label{sec:obs_line_strength}
\begin{longrotatetable}
\startlongtable
\begin{deluxetable*}{lcccccccccccccccc}
\tablecolumns{9}
\setlength{\tabcolsep}{5pt}
\tablecaption{RM AGN UV+optical spectroscopic properties}\label{tab:UV_spec_prop}
\tabletypesize{\scriptsize}
\tablehead{
    \colhead{Objects } &
    \colhead{$\log{\L5100}$ } &
    \colhead{$\log{\lambda L_{1350}}$ } &
    \colhead{$\log{L_{\rm CIV}}$ } &
    \colhead{REW(CIV) } &
    \colhead{CIV blueshift } &
    \colhead{$\log{L_{\rm CIII]}}$ } &
    \colhead{ REW(CIII]) } &
    \colhead{$\log{L_{\rm HeII1640}}$ } &
    \colhead{ REW(HeII1640) } &
    \colhead{$\log{L_{\rm MgII}}$ } &
    \colhead{REW(MgII) } &
    \colhead{$\log{L_{\rm H\beta}}$ } &
    \colhead{REW(H$\beta$) } &\\
    \colhead{} &
    \colhead{\ergs} &
    \colhead{\ergs} &
    \colhead{\ergs} &
    \colhead{$\rm \AA$} &
    \colhead{\kms} &
    \colhead{\ergs} &
    \colhead{$\rm \AA$} &
    \colhead{\ergs} &
    \colhead{$\rm \AA$} &
    \colhead{\ergs} &
    \colhead{$\rm \AA$} &
    \colhead{\ergs} &
    \colhead{$\rm \AA$} &
}
\startdata
RM160 & $43.79 \pm 0.01$ & $44.34 \pm 0.02$ &  $43.62 \pm 0.01$ &  $321.2 \pm 5.7$ &  $-595 \pm 45$ &  $42.47 \pm 0.03$ &  $33.0 \pm 2.5$ &  $42.19 \pm 0.09$ &  $13.0 \pm 2.7$ &  ... & ... & $42.07 \pm 0.01$ &  $96.5 \pm 1.4$ & $40.9 \pm 0.06$ &  $6.4 \pm 0.6$ \\ 
RM177 & $43.97 \pm 0.01$ & $44.08 \pm 0.03$ &  $42.64 \pm 0.03$ &  $47.7 \pm 3.9$ &  $145 \pm 366$ &  $42.32 \pm 0.09$ &  $29.5 \pm 5.7$ &  $41.37 \pm 0.23$ &  $2.7 \pm 2.5$ &  ... & ... & $42.31 \pm 0.01$ &  $62.0 \pm 1.7$ & $41.4 \pm 0.02$ &  $7.5 \pm 0.4$ \\ 
RM229 & $43.56 \pm 0.01$ & $44.25 \pm 0.03$ &  $43.0 \pm 0.05$ &  $100.1 \pm 9.1$ &  $198 \pm 504$ &  $42.35 \pm 0.27$ &  $36.4 \pm 16.5$ &  $41.83 \pm 0.24$ &  $7.8 \pm 6.5$ &  ... & ... & $41.93 \pm 0.02$ &  $62.0 \pm 2.3$ & $40.86 \pm 0.15$ &  $5.2 \pm 2.5$ \\ 
RM267 & $44.11 \pm 0.01$ & $44.76 \pm 0.02$ &  $43.26 \pm 0.02$ &  $46.7 \pm 2.1$ &  $332 \pm 166$ &  $42.94 \pm 0.12$ &  $47.6 \pm 14.0$ &  ... & ... &  ... & ... & $42.54 \pm 0.02$ &  $81.1 \pm 4.2$ & $41.33 \pm 0.17$ &  $4.8 \pm 2.5$ \\ 
RM272 & $43.94 \pm 0.01$ & $44.26 \pm 0.05$ &  $43.16 \pm 0.04$ &  $145.4 \pm 15.3$ &  $-287 \pm 127$ &  $42.24 \pm 0.05$ &  $21.0 \pm 2.6$ &  $42.2 \pm 0.32$ &  $17.1 \pm 9.5$ &  ... & ... & $42.37 \pm 0.01$ &  $132.2 \pm 0.6$ & $41.46 \pm 0.01$ &  $15.8 \pm 0.5$ \\ 
RM301 & $44.04 \pm 0.01$ & $43.75 \pm 0.09$ &  $42.82 \pm 0.03$ &  $183.9 \pm 12.4$ &  $304 \pm 193$ &  $41.83 \pm 0.23$ &  $20.4 \pm 11.7$ &  ... & ... &  ... & ... & $42.27 \pm 0.01$ &  $58.7 \pm 1.2$ & $40.75 \pm 0.08$ &  $1.7 \pm 0.3$ \\ 
RM305 & $44.19 \pm 0.01$ & $44.37 \pm 0.04$ &  $42.91 \pm 0.04$ &  $54.5 \pm 5.2$ &  $-1210 \pm 881$ &  $42.61 \pm 0.18$ &  $40.5 \pm 17.8$ &  $41.47 \pm 0.33$ &  $2.1 \pm 2.4$ &  ... & ... & $42.53 \pm 0.04$ &  $77.7 \pm 6.2$ & $41.4 \pm 0.02$ &  $5.6 \pm 0.3$ \\ 
RM320 & $43.44 \pm 0.01$ & $43.59 \pm 0.06$ &  $42.68 \pm 0.04$ &  $138.7 \pm 13.4$ &  $407 \pm 266$ &  $41.92 \pm 0.05$ &  $25.5 \pm 2.9$ &  $42.02 \pm 0.26$ &  $29.9 \pm 13.2$ &  ... & ... & $41.89 \pm 0.01$ &  $74.9 \pm 1.9$ & $41.18 \pm 0.12$ &  $14.2 \pm 4.7$ \\ 
RM371 & $44.09 \pm 0.01$ & $44.03 \pm 0.1$ &  $43.17 \pm 0.04$ &  $205.4 \pm 24.3$ &  $1165 \pm 435$ &  $42.42 \pm 0.28$ &  $37.0 \pm 13.8$ &  $42.27 \pm 0.11$ &  $25.4 \pm 6.3$ &  ... & ... & $42.4 \pm 0.01$ &  $101.6 \pm 2.0$ & $40.66 \pm 0.14$ &  $1.8 \pm 0.6$ \\ 
RM622 & $44.3 \pm 0.01$ & $44.49 \pm 0.03$ &  $43.03 \pm 0.07$ &  $49.8 \pm 7.5$ &  $-1222 \pm 609$ &  $42.54 \pm 0.24$ &  $20.8 \pm 9.9$ &  $41.85 \pm 0.26$ &  $3.5 \pm 3.4$ &  ... & ... & $42.62 \pm 0.01$ &  $81.1 \pm 1.1$ & $41.14 \pm 0.04$ &  $2.6 \pm 0.3$ \\ 
RM720 & $44.3 \pm 0.01$ & $44.22 \pm 0.04$ &  $43.1 \pm 0.07$ &  $114.3 \pm 16.7$ &  $-328 \pm 240$ &  $42.29 \pm 0.14$ &  $20.4 \pm 5.2$ &  $42.29 \pm 0.09$ &  $18.0 \pm 3.8$ &  ... & ... & $42.46 \pm 0.01$ &  $61.8 \pm 0.4$ & $41.5 \pm 0.03$ &  $6.4 \pm 0.5$ \\ 
RM781 & $43.6 \pm 0.01$ & $43.88 \pm 0.09$ &  $42.49 \pm 0.05$ &  $55.1 \pm 5.9$ &  $-189 \pm 343$ &  $41.63 \pm 0.15$ &  $10.0 \pm 4.0$ &  $41.36 \pm 0.24$ &  $4.2 \pm 4.1$ &  ... & ... & $41.91 \pm 0.02$ &  $65.8 \pm 2.8$ & $41.27 \pm 0.05$ &  $14.6 \pm 1.6$ \\ 
RM782 & $43.93 \pm 0.01$ & $43.04 \pm 0.31$ &  $42.82 \pm 0.04$ &  $66.3 \pm 5.9$ &  $-586 \pm 431$ &  $41.8 \pm 0.11$ &  $6.0 \pm 1.5$ &  ... & ... &  ... & ... & $42.0 \pm 0.01$ &  $37.0 \pm 0.5$ & $40.5 \pm 0.06$ &  $1.1 \pm 0.2$ \\ 
RM790 & $43.33 \pm 0.01$ & ... &  ... & ... & ... &  $41.29 \pm 0.18$ &  $13.4 \pm 4.7$ &  $41.16 \pm 0.3$ &  $20.2 \pm 15.1$ &  ... & ... & $41.87 \pm 0.03$ &  $53.9 \pm 3.2$ & $24.03 \pm 8.55$ &  $0.0 \pm 5.0$ \\ 
RM840 & $43.24 \pm 0.01$ & $43.66 \pm 0.13$ &  $42.65 \pm 0.05$ &  $153.7 \pm 17.4$ &  $-218 \pm 269$ &  $42.07 \pm 0.07$ &  $53.3 \pm 7.3$ &  $41.47 \pm 0.31$ &  $10.8 \pm 5.2$ &  ... & ... & $41.6 \pm 0.01$ &  $38.9 \pm 0.7$ & $39.47 \pm 19.82$ &  $0.3 \pm 0.2$ \\ 
Mrk335 & $43.76 \pm 0.07$ & $43.72 \pm 0.01$ &  $42.63 \pm 0.01$ &  $108.2 \pm 1.8$ &  $-127 \pm 23$ &  $42.8 \pm 0.02$ &  $135.6 \pm 7.5$ &  $41.5 \pm 0.02$ &  $7.9 \pm 0.3$ &  $41.91 \pm 0.03$ &  $15.5 \pm 1.0$ & $42.09 \pm 0.09$ &  $108.2 \pm 0.1$ &  ... & ... \\ 
PG0026 & $44.97 \pm 0.02$ & $45.26 \pm 0.01$ &  $43.55 \pm 0.01$ &  $32.8 \pm 1.1$ &  $-993 \pm 129$ &  ... & ... &  $42.8 \pm 0.15$ &  $6.5 \pm 3.7$ &  ... & ... & $42.93 \pm 0.04$ &  $46.2 \pm 0.0$ &  ... & ... \\ 
PG0052 & $44.81 \pm 0.03$ & $45.36 \pm 0.01$ &  $44.18 \pm 0.03$ &  $113.7 \pm 7.6$ &  $670 \pm 51$ &  $42.53 \pm 0.18$ &  $3.8 \pm 2.4$ &  $42.47 \pm 0.61$ &  $2.5 \pm 9.0$ &  ... & ... & $43.13 \pm 0.05$ &  $107.4 \pm 0.0$ &  ... & ... \\ 
Fairall9 & $43.98 \pm 0.04$ & $44.43 \pm 0.01$ &  $43.33 \pm 0.01$ &  $114.9 \pm 0.9$ &  $61 \pm 24$ &  $42.68 \pm 0.12$ &  $29.0 \pm 10.2$ &  $42.53 \pm 0.18$ &  $19.1 \pm 5.4$ &  $42.84 \pm 0.01$ &  $53.5 \pm 0.5$ & $42.67 \pm 0.04$ &  $249.8 \pm 0.0$ &  ... & ... \\ 
Mrk590 & $43.5 \pm 0.21$ & $42.92 \pm 0.01$ &  $42.27 \pm 0.01$ &  $710.5 \pm 7.3$ &  $356 \pm 89$ &  $40.18 \pm 0.3$ &  $2.4 \pm 9.4$ &  $40.92 \pm 0.09$ &  $32.4 \pm 7.0$ &  $41.92 \pm 0.01$ &  $48.1 \pm 0.8$ & $41.85 \pm 0.12$ &  $108.6 \pm 0.1$ &  ... & ... \\ 
Mrk1044 & $43.1 \pm 0.1$ & $43.42 \pm 0.01$ &  $41.85 \pm 0.01$ &  $42.7 \pm 0.3$ &  $865 \pm 26$ &  $41.08 \pm 0.21$ &  $10.0 \pm 7.1$ &  $41.58 \pm 0.01$ &  $24.8 \pm 0.7$ &  $40.93 \pm 0.06$ &  $10.9 \pm 1.5$ & $41.39 \pm 0.09$ &  $101.4 \pm 0.1$ &  ... & ... \\ 
3C120 & $44.0 \pm 0.1$ & $44.41 \pm 0.01$ &  $43.31 \pm 0.01$ &  $134.9 \pm 1.5$ &  $13 \pm 38$ &  $42.04 \pm 0.1$ &  $10.5 \pm 2.4$ &  $42.39 \pm 0.02$ &  $17.7 \pm 0.8$ &  $42.67 \pm 0.01$ &  $72.3 \pm 1.3$ & $42.36 \pm 0.04$ &  $118.8 \pm 0.0$ &  ... & ... \\ 
Ark120 & $43.87 \pm 0.25$ & $44.61 \pm 0.01$ &  $43.32 \pm 0.01$ &  $77.9 \pm 1.0$ &  $20 \pm 25$ &  $42.91 \pm 0.04$ &  $36.2 \pm 3.3$ &  $42.69 \pm 0.12$ &  $18.8 \pm 4.1$ &  $42.76 \pm 0.01$ &  $37.5 \pm 0.4$ & $42.54 \pm 0.13$ &  $244.8 \pm 0.1$ &  ... & ... \\ 
MCG811 & $43.33 \pm 0.11$ & $42.81 \pm 0.04$ &  $41.91 \pm 0.01$ &  $172.7 \pm 5.5$ &  $643 \pm 170$ &  $41.54 \pm 0.16$ &  $45.5 \pm 13.7$ &  $41.15 \pm 0.04$ &  $27.2 \pm 2.8$ &  $41.96 \pm 0.01$ &  $72.3 \pm 1.1$ & $41.66 \pm 0.09$ &  $108.7 \pm 0.1$ &  ... & ... \\ 
Mrk374 & $43.77 \pm 0.04$ & $43.72 \pm 0.02$ &  $42.8 \pm 0.06$ &  $206.6 \pm 26.5$ &  $-20 \pm 227$ &  $41.76 \pm 0.11$ &  $24.9 \pm 7.5$ &  $40.56 \pm 0.63$ &  $1.3 \pm 9.5$ &  $41.91 \pm 0.04$ &  $43.1 \pm 3.5$ & $41.83 \pm 0.04$ &  $58.3 \pm 0.0$ &  ... & ... \\ 
Mrk79 & $43.68 \pm 0.07$ & $43.59 \pm 0.01$ &  $42.46 \pm 0.01$ &  $115.7 \pm 3.7$ &  $2446 \pm 75$ &  $42.0 \pm 0.05$ &  $46.9 \pm 4.7$ &  $42.06 \pm 0.05$ &  $49.0 \pm 4.5$ &  $41.75 \pm 0.02$ &  $26.2 \pm 1.5$ & $41.9 \pm 0.05$ &  $85.4 \pm 0.0$ &  ... & ... \\ 
Mrk382 & $43.12 \pm 0.08$ & $43.15 \pm 0.01$ &  ... & ... & ... &  ... & ... &  ... & ... &  $40.15 \pm 0.18$ &  $16.4 \pm 7.1$ & $41.01 \pm 0.05$ &  $39.6 \pm 0.0$ &  ... & ... \\ 
PG0804 & $44.91 \pm 0.02$ & $45.36 \pm 0.01$ &  $44.04 \pm 0.01$ &  $87.5 \pm 1.4$ &  $-923 \pm 596$ &  $42.66 \pm 0.19$ &  $6.3 \pm 4.1$ &  $42.78 \pm 0.06$ &  $5.3 \pm 0.7$ &  $43.0 \pm 0.04$ &  $19.2 \pm 1.7$ & $43.29 \pm 0.03$ &  $122.5 \pm 0.0$ &  ... & ... \\ 
PG0844 & $44.22 \pm 0.07$ & $44.59 \pm 0.01$ &  $43.01 \pm 0.02$ &  $38.6 \pm 1.5$ &  $-141 \pm 59$ &  $42.8 \pm 0.18$ &  $28.3 \pm 8.9$ &  $42.67 \pm 0.02$ &  $18.6 \pm 1.0$ &  $42.27 \pm 0.08$ &  $11.3 \pm 2.4$ & $42.56 \pm 0.05$ &  $111.2 \pm 0.0$ &  ... & ... \\ 
Mrk110 & $43.66 \pm 0.12$ & $43.92 \pm 0.01$ &  $43.05 \pm 0.01$ &  $186.4 \pm 2.1$ &  $-762 \pm 37$ &  $41.92 \pm 0.08$ &  $14.5 \pm 2.7$ &  $42.02 \pm 0.01$ &  $17.3 \pm 0.6$ &  $42.4 \pm 0.01$ &  $65.2 \pm 0.5$ & $42.03 \pm 0.08$ &  $123.8 \pm 0.1$ &  ... & ... \\ 
PG0953 & $45.19 \pm 0.01$ & $45.79 \pm 0.01$ &  $44.71 \pm 0.02$ &  $334.5 \pm 17.5$ &  $52 \pm 52$ &  ... & ... &  ... & ... &  ... & ... & $43.29 \pm 0.04$ &  $64.7 \pm 0.0$ &  ... & ... \\ 
NGC3227 & $42.24 \pm 0.11$ & $39.99 \pm 0.12$ &  ... & ... & ... &  $39.74 \pm 0.1$ &  $50.8 \pm 9.4$ &  $38.66 \pm 0.15$ &  $18.5 \pm 9.3$ &  $40.17 \pm 0.01$ &  $38.7 \pm 1.0$ & $40.38 \pm 0.1$ &  $71.0 \pm 0.1$ &  ... & ... \\ 
NGC3516 & $42.79 \pm 0.2$ & $43.11 \pm 0.01$ &  $41.8 \pm 0.01$ &  $69.3 \pm 0.4$ &  $-1711 \pm 73$ &  $41.74 \pm 0.12$ &  $71.4 \pm 14.8$ &  $41.04 \pm 0.01$ &  $12.3 \pm 0.3$ &  $41.5 \pm 0.02$ &  $77.4 \pm 2.8$ & $41.06 \pm 0.18$ &  $94.7 \pm 0.2$ &  ... & ... \\ 
SBS1116 & $42.14 \pm 0.23$ & $42.85 \pm 0.01$ &  $41.95 \pm 0.02$ &  $231.6 \pm 13.5$ &  $-220 \pm 62$ &  $41.12 \pm 0.04$ &  $39.8 \pm 3.8$ &  $40.72 \pm 0.07$ &  $14.7 \pm 2.6$ &  $41.02 \pm 0.04$ &  $38.2 \pm 3.4$ & $40.7 \pm 0.07$ &  $186.8 \pm 0.1$ &  ... & ... \\ 
Arp151 & $42.55 \pm 0.1$ & $42.6 \pm 0.01$ &  $41.76 \pm 0.02$ &  $233.0 \pm 8.6$ &  $-2167 \pm 132$ &  $41.02 \pm 0.05$ &  $52.1 \pm 5.5$ &  $40.39 \pm 0.06$ &  $10.6 \pm 1.5$ &  $40.89 \pm 0.01$ &  $47.3 \pm 0.8$ & $40.95 \pm 0.11$ &  $130.0 \pm 0.1$ &  ... & ... \\ 
NGC3783 & $42.56 \pm 0.18$ & $43.34 \pm 0.01$ &  $42.4 \pm 0.01$ &  $171.1 \pm 2.5$ &  $-137 \pm 29$ &  $41.7 \pm 0.1$ &  $32.7 \pm 7.1$ &  $41.29 \pm 0.02$ &  $13.4 \pm 0.6$ &  $41.83 \pm 0.01$ &  $54.9 \pm 1.2$ & $41.01 \pm 0.18$ &  $144.0 \pm 0.2$ &  ... & ... \\ 
Mrk1310 & $42.29 \pm 0.14$ & $41.76 \pm 0.01$ &  $41.23 \pm 0.03$ &  $485.7 \pm 37.6$ &  $-183 \pm 250$ &  $40.1 \pm 0.27$ &  $28.7 \pm 31.6$ &  ... & ... &  $40.65 \pm 0.06$ &  $64.7 \pm 9.4$ & $40.56 \pm 0.1$ &  $94.3 \pm 0.1$ &  ... & ... \\ 
NGC4051 & $41.9 \pm 0.15$ & $41.36 \pm 0.01$ &  $39.99 \pm 0.01$ &  $62.1 \pm 1.4$ &  $482 \pm 516$ &  $39.79 \pm 0.1$ &  $35.7 \pm 7.5$ &  $39.35 \pm 0.02$ &  $13.9 \pm 0.7$ &  $40.03 \pm 0.01$ &  $61.1 \pm 1.8$ & $40.09 \pm 0.19$ &  $78.6 \pm 0.2$ &  ... & ... \\ 
NGC4151 & $42.09 \pm 0.21$ & $42.9 \pm 0.02$ &  $42.12 \pm 0.01$ &  $210.0 \pm 2.6$ &  $139 \pm 44$ &  $41.36 \pm 0.03$ &  $37.3 \pm 2.4$ &  $40.76 \pm 0.02$ &  $8.9 \pm 0.4$ &  $41.37 \pm 0.01$ &  $46.3 \pm 0.7$ & $40.56 \pm 0.2$ &  $150.8 \pm 0.2$ &  ... & ... \\ 
PG1211 & $44.73 \pm 0.08$ & $44.9 \pm 0.01$ &  $43.74 \pm 0.02$ &  $111.8 \pm 5.5$ &  $499 \pm 28$ &  $43.04 \pm 0.07$ &  $28.8 \pm 4.5$ &  $42.55 \pm 0.04$ &  $7.6 \pm 0.7$ &  $42.38 \pm 0.06$ &  $7.3 \pm 1.0$ & $43.02 \pm 0.06$ &  $100.2 \pm 0.1$ &  ... & ... \\ 
NGC4253 & $42.57 \pm 0.12$ & $41.9 \pm 0.03$ &  $40.48 \pm 0.06$ &  $61.6 \pm 7.9$ &  $35 \pm 267$ &  $39.6 \pm 0.25$ &  $8.2 \pm 11.3$ &  ... & ... &  $40.41 \pm 0.08$ &  $30.7 \pm 5.8$ & $40.77 \pm 0.12$ &  $81.1 \pm 0.1$ &  ... & ... \\ 
PG1229 & $43.7 \pm 0.05$ & $44.56 \pm 0.01$ &  $43.19 \pm 0.01$ &  $62.3 \pm 2.0$ &  $1038 \pm 183$ &  $42.78 \pm 0.04$ &  $29.8 \pm 2.3$ &  $42.74 \pm 0.12$ &  $22.9 \pm 4.9$ &  $42.52 \pm 0.02$ &  $27.7 \pm 1.5$ & $42.31 \pm 0.06$ &  $209.7 \pm 0.1$ &  ... & ... \\ 
NGC4593 & $42.62 \pm 0.37$ & $42.36 \pm 0.01$ &  $41.5 \pm 0.01$ &  $198.8 \pm 1.2$ &  $-113 \pm 32$ &  $40.66 \pm 0.21$ &  $21.6 \pm 11.8$ &  $40.42 \pm 0.24$ &  $15.4 \pm 14.0$ &  $41.14 \pm 0.01$ &  $61.7 \pm 0.6$ & $40.95 \pm 0.33$ &  $108.3 \pm 0.3$ &  ... & ... \\ 
NGC4748 & $42.56 \pm 0.12$ & $42.87 \pm 0.01$ &  $41.52 \pm 0.02$ &  $70.4 \pm 3.5$ &  $-1069 \pm 559$ &  $41.66 \pm 0.07$ &  $110.9 \pm 15.9$ &  $41.22 \pm 0.08$ &  $36.5 \pm 6.6$ &  $40.99 \pm 0.01$ &  $31.3 \pm 0.9$ & $40.98 \pm 0.1$ &  $136.8 \pm 0.1$ &  ... & ... \\ 
PG1307 & $44.85 \pm 0.02$ & $45.34 \pm 0.01$ &  $44.06 \pm 0.02$ &  $90.0 \pm 3.9$ &  $975 \pm 56$ &  ... & ... &  ... & ... &  $42.6 \pm 0.12$ &  $6.8 \pm 1.8$ & $43.13 \pm 0.06$ &  $98.4 \pm 0.1$ &  ... & ... \\ 
MCG630 & $41.64 \pm 0.11$ & $40.25 \pm 0.37$ &  ... & ... & ... &  $39.92 \pm 0.21$ &  $80.6 \pm 45.2$ &  ... & ... &  ... & ... & $39.85 \pm 0.19$ &  $91.0 \pm 0.2$ &  ... & ... \\ 
NGC5273 & $41.54 \pm 0.16$ & $41.0 \pm 0.05$ &  $39.89 \pm 0.1$ &  $175.2 \pm 37.2$ &  $-352 \pm 736$ &  $39.62 \pm 0.29$ &  $277.1 \pm 131.1$ &  $39.29 \pm 0.27$ &  $54.9 \pm 22.4$ &  $38.54 \pm 0.16$ &  $9.6 \pm 3.6$ & $39.74 \pm 0.11$ &  $82.2 \pm 0.1$ &  ... & ... \\ 
Mrk279 & $43.71 \pm 0.07$ & $43.99 \pm 0.01$ &  $41.76 \pm 0.02$ &  $9.7 \pm 0.5$ &  $1160 \pm 184$ &  $41.79 \pm 0.23$ &  $13.7 \pm 12.7$ &  $41.53 \pm 0.02$ &  $6.1 \pm 0.3$ &  ... & ... & $42.12 \pm 0.06$ &  $132.2 \pm 0.1$ &  ... & ... \\ 
PG1411 & $44.56 \pm 0.02$ & $44.65 \pm 0.01$ &  $43.33 \pm 0.02$ &  $72.1 \pm 2.4$ &  $-1619 \pm 77$ &  $43.15 \pm 0.35$ &  $57.8 \pm 33.8$ &  $42.31 \pm 0.03$ &  $7.2 \pm 0.5$ &  ... & ... & $42.85 \pm 0.03$ &  $99.7 \pm 0.0$ &  ... & ... \\ 
NGC5548 & $43.3 \pm 0.19$ & $43.58 \pm 0.01$ &  $42.55 \pm 0.01$ &  $134.3 \pm 0.5$ &  $549 \pm 28$ &  $42.28 \pm 0.03$ &  $71.2 \pm 4.9$ &  $41.76 \pm 0.01$ &  $21.7 \pm 0.4$ &  $42.17 \pm 0.01$ &  $74.5 \pm 0.4$ & $41.65 \pm 0.21$ &  $117.8 \pm 0.2$ &  ... & ... \\ 
PG1426 & $44.63 \pm 0.02$ & $45.2 \pm 0.01$ &  $43.72 \pm 0.05$ &  $55.6 \pm 6.0$ &  $208 \pm 215$ &  $43.13 \pm 0.05$ &  $20.3 \pm 2.4$ &  $43.2 \pm 0.05$ &  $18.2 \pm 2.2$ &  $42.97 \pm 0.04$ &  $21.9 \pm 2.1$ & $42.83 \pm 0.04$ &  $80.1 \pm 0.0$ &  ... & ... \\ 
Mrk817 & $43.74 \pm 0.09$ & $44.29 \pm 0.01$ &  $42.82 \pm 0.01$ &  $58.1 \pm 0.7$ &  $357 \pm 38$ &  $41.72 \pm 0.25$ &  $7.0 \pm 4.0$ &  $42.25 \pm 0.01$ &  $17.4 \pm 0.3$ &  $42.37 \pm 0.05$ &  $58.9 \pm 6.0$ & $41.93 \pm 0.14$ &  $78.5 \pm 0.1$ &  ... & ... \\ 
Mrk1511 & $43.16 \pm 0.06$ & $43.48 \pm 0.01$ &  $42.32 \pm 0.03$ &  $112.0 \pm 8.3$ &  $-245 \pm 238$ &  $41.37 \pm 0.15$ &  $15.4 \pm 4.9$ &  ... & ... &  $41.56 \pm 0.09$ &  $25.9 \pm 6.5$ & $41.52 \pm 0.06$ &  $115.5 \pm 0.1$ &  ... & ... \\ 
Mrk290 & $43.17 \pm 0.06$ & $43.77 \pm 0.01$ &  $42.71 \pm 0.01$ &  $129.0 \pm 1.8$ &  $321 \pm 45$ &  $42.15 \pm 0.13$ &  $39.0 \pm 9.9$ &  $41.86 \pm 0.04$ &  $18.3 \pm 1.5$ &  $42.1 \pm 0.01$ &  $54.9 \pm 0.9$ & $41.64 \pm 0.06$ &  $153.0 \pm 0.1$ &  ... & ... \\ 
Mrk486 & $43.69 \pm 0.05$ & $43.17 \pm 0.02$ &  $41.88 \pm 0.16$ &  $55.2 \pm 15.3$ &  $-1110 \pm 70$ &  $41.0 \pm 0.21$ &  $5.5 \pm 3.7$ &  $41.16 \pm 0.09$ &  $9.7 \pm 2.1$ &  $42.01 \pm 0.07$ &  $41.6 \pm 7.2$ & $42.12 \pm 0.04$ &  $135.9 \pm 0.0$ &  ... & ... \\ 
PG1613 & $44.77 \pm 0.02$ & $45.18 \pm 0.01$ &  $43.93 \pm 0.01$ &  $88.7 \pm 1.3$ &  $1117 \pm 84$ &  $42.8 \pm 0.18$ &  $9.6 \pm 4.5$ &  ... & ... &  $43.35 \pm 0.02$ &  $49.6 \pm 2.4$ & $43.0 \pm 0.03$ &  $86.7 \pm 0.0$ &  ... & ... \\ 
PG1617 & $44.39 \pm 0.02$ & $44.87 \pm 0.01$ &  $42.98 \pm 0.02$ &  $21.5 \pm 0.9$ &  $-72 \pm 123$ &  $43.26 \pm 0.04$ &  $53.5 \pm 4.7$ &  $41.91 \pm 0.07$ &  $2.0 \pm 0.3$ &  $43.2 \pm 0.03$ &  $69.8 \pm 5.1$ & $42.74 \pm 0.05$ &  $114.8 \pm 0.0$ &  ... & ... \\ 
3C382 & $43.84 \pm 0.1$ & $44.48 \pm 0.01$ &  $43.52 \pm 0.01$ &  $170.3 \pm 5.0$ &  $39 \pm 51$ &  $42.21 \pm 0.13$ &  $11.1 \pm 2.9$ &  $41.92 \pm 0.29$ &  $4.6 \pm 5.2$ &  $42.8 \pm 0.01$ &  $64.3 \pm 1.2$ & $42.54 \pm 0.03$ &  $259.3 \pm 0.0$ &  ... & ... \\ 
3C390.3 & $44.43 \pm 0.58$ & $43.73 \pm 0.01$ &  $42.79 \pm 0.01$ &  $226.6 \pm 3.4$ &  $1474 \pm 84$ &  $42.4 \pm 0.2$ &  $112.6 \pm 33.4$ &  $42.1 \pm 0.23$ &  $52.5 \pm 17.9$ &  $42.64 \pm 0.01$ &  $113.5 \pm 3.3$ & $42.6 \pm 0.35$ &  $108.8 \pm 0.4$ &  ... & ... \\ 
NGC6814 & $42.12 \pm 0.28$ & $41.66 \pm 0.02$ &  $40.61 \pm 0.03$ &  $182.6 \pm 11.7$ &  $-167 \pm 416$ &  $40.31 \pm 0.14$ &  $132.6 \pm 39.1$ &  $38.95 \pm 0.25$ &  $4.6 \pm 2.0$ &  $40.29 \pm 0.01$ &  $87.3 \pm 2.3$ & $40.5 \pm 0.28$ &  $121.6 \pm 0.3$ &  ... & ... \\ 
Mrk509 & $44.19 \pm 0.05$ & $44.62 \pm 0.01$ &  $43.49 \pm 0.01$ &  $116.7 \pm 0.9$ &  $344 \pm 31$ &  $43.12 \pm 0.03$ &  $65.2 \pm 4.7$ &  $42.55 \pm 0.05$ &  $14.5 \pm 1.6$ &  $42.86 \pm 0.01$ &  $49.7 \pm 0.7$ & $42.61 \pm 0.04$ &  $132.7 \pm 0.0$ &  ... & ... \\ 
NGC7469 & $43.51 \pm 0.11$ & $43.85 \pm 0.01$ &  $42.49 \pm 0.01$ &  $66.7 \pm 0.6$ &  $-253 \pm 40$ &  $41.79 \pm 0.06$ &  $16.4 \pm 2.1$ &  $41.93 \pm 0.02$ &  $19.3 \pm 1.1$ &  $42.1 \pm 0.01$ &  $50.2 \pm 1.7$ & $41.6 \pm 0.1$ &  $63.0 \pm 0.1$ &  ... & ... \\ 
Zw229 & $42.71 \pm 0.05$ & $43.16 \pm 0.01$ &  $42.25 \pm 0.04$ &  $187.0 \pm 20.8$ &  $870 \pm 485$ &  $41.37 \pm 0.04$ &  $28.1 \pm 2.7$ &  $41.16 \pm 0.37$ &  $16.1 \pm 13.8$ &  $41.52 \pm 0.01$ &  $49.6 \pm 1.6$ & ... & ... & ... & ... \\ 
Mrk1501 & $44.32 \pm 0.05$ & $44.7 \pm 0.01$ &  $43.61 \pm 0.01$ &  $120.3 \pm 3.0$ &  $-108 \pm 88$ &  $43.18 \pm 0.13$ &  $48.6 \pm 10.9$ &  $43.08 \pm 0.07$ &  $36.0 \pm 4.8$ &  $42.99 \pm 0.05$ &  $35.0 \pm 3.6$ & ... & ... & ... & ... \\ 
Mrk50 & $42.88 \pm 0.07$ & $43.22 \pm 0.01$ &  $42.16 \pm 0.02$ &  $152.6 \pm 6.4$ &  $-57 \pm 86$ &  $41.01 \pm 0.16$ &  $14.4 \pm 7.6$ &  ... & ... &  $41.4 \pm 0.01$ &  $45.4 \pm 0.7$ & ... & ... & ... & ... \\ 
NGC4395 & $39.77 \pm 0.02$ & $39.82 \pm 0.01$ &  $38.62 \pm 0.05$ &  $82.9 \pm 11.5$ &  $1187 \pm 544$ &  ... & ... &  $37.83 \pm 0.02$ &  $14.0 \pm 0.6$ &  ... & ... & ... & ... & ... & ... \\ 
Mrk6 & $43.75 \pm 0.06$ & $42.09 \pm 0.1$ &  $40.9 \pm 0.16$ &  $91.0 \pm 29.6$ &  $4497 \pm 2534$ &  $40.33 \pm 0.18$ &  $22.5 \pm 10.5$ &  $40.44 \pm 0.08$ &  $31.3 \pm 5.5$ &  $40.94 \pm 0.04$ &  $64.7 \pm 6.3$ & ... & ... & ... & ... 
\enddata
\end{deluxetable*}
\end{longrotatetable}

\begin{figure*}\centering
    \includegraphics[width=0.8\linewidth]{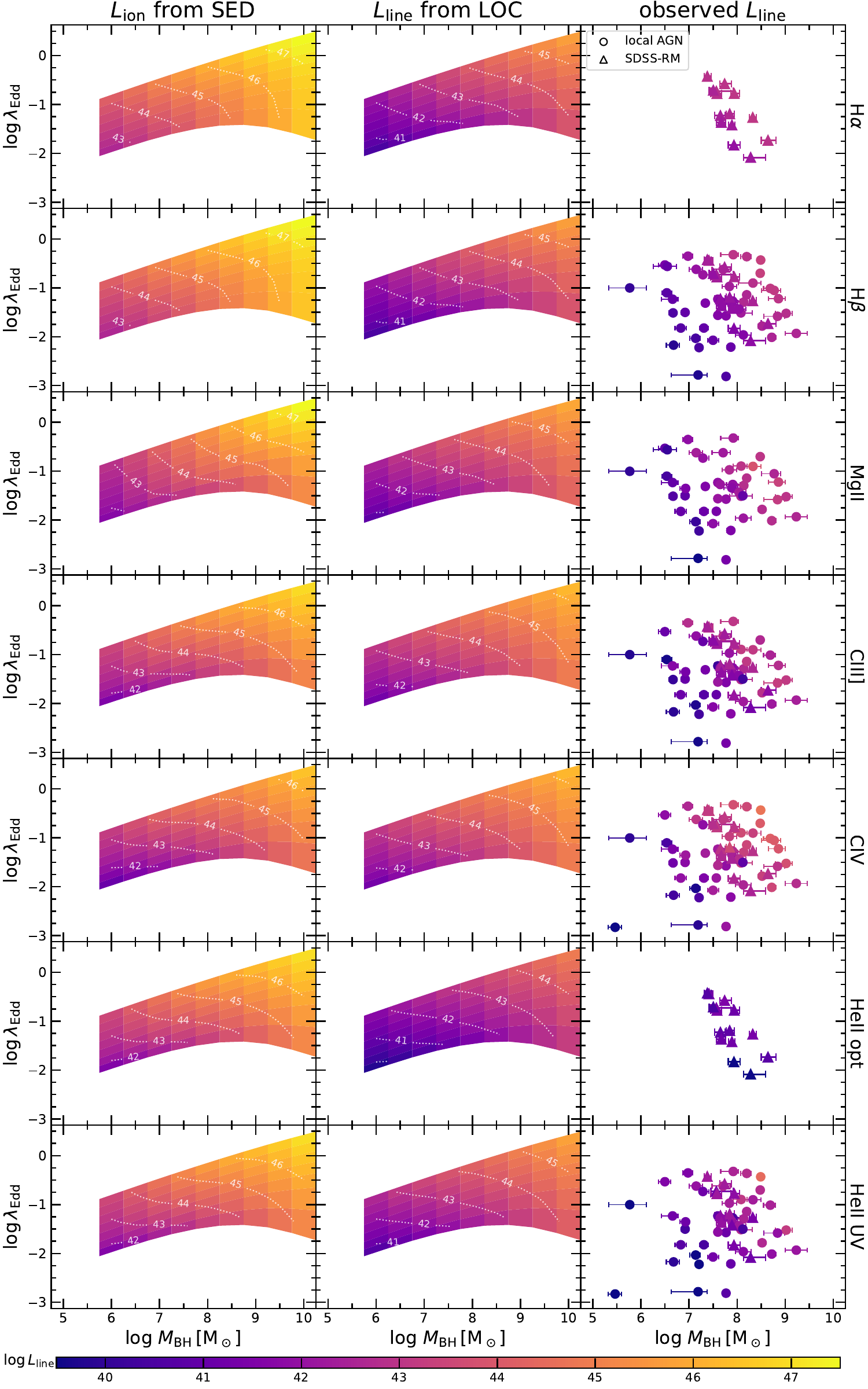}
    \caption{Distributions of line luminosities in \Mbh-\Redd\ plane. The left panel shows the integrated ionizing luminosity ($L_{\rm ion}=\int^{\nu_{\rm destruct}}_{\nu_{\rm product}}L_\nu\,d\nu$ within the production and destruction photon energies); the middle panel displays the line luminosities predicted by the LOC model; and the right panel presents the observed line luminosities as a visual assessment of the theoretical photoionization. The positive correlations in all panels demonstrate the consistency between our theoretical computations and observations. 
    } 
    \label{fig:Lflux_grid}
\end{figure*}

\begin{figure*}\centering
    \includegraphics[width=0.8\linewidth]{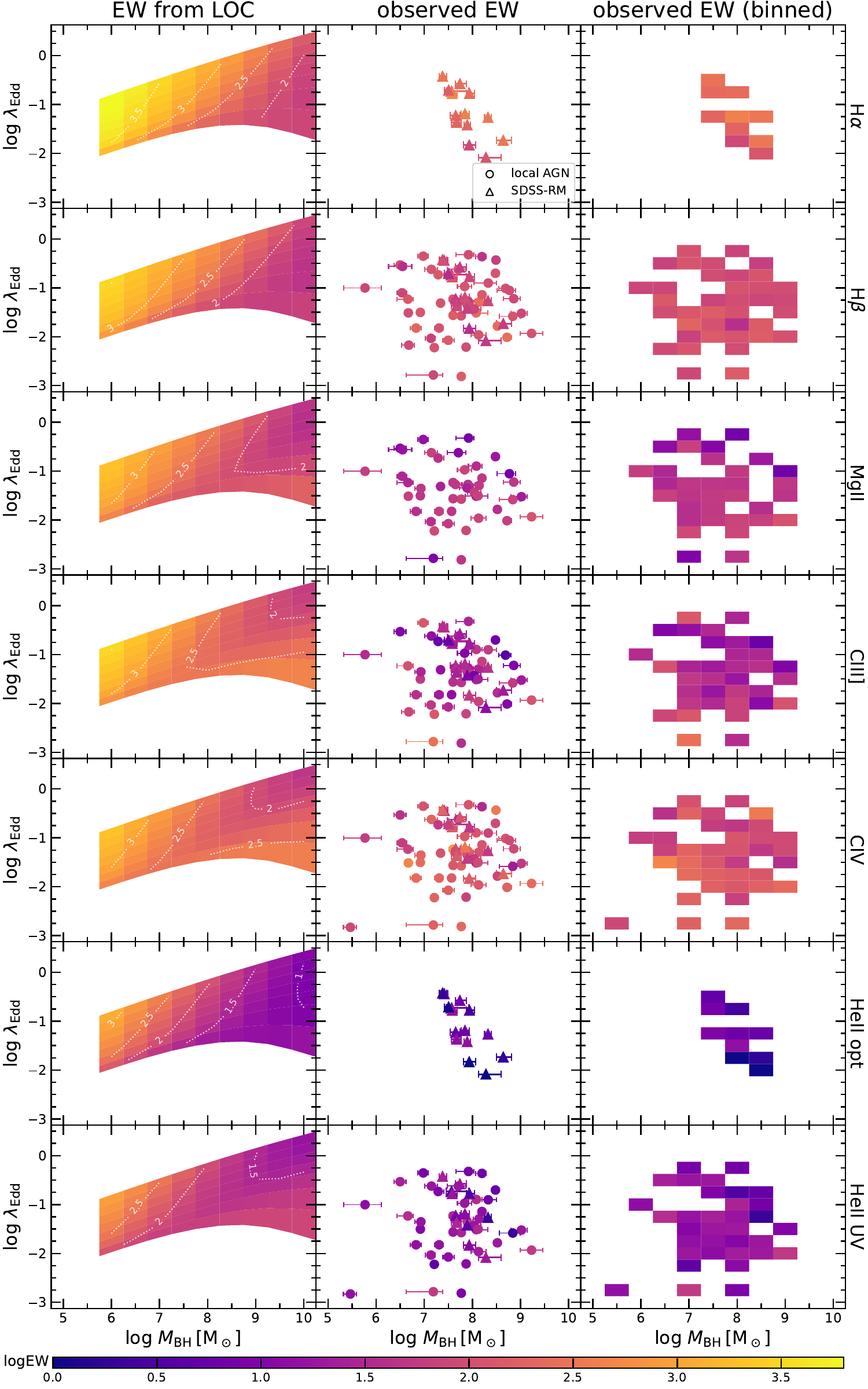}
    \caption{Distributions of emission line equivalent width (EW) in the \Mbh-\Redd\ plane. The left panel displays the EW distributions predicted by the LOC model. The middle panel presents the observed EW from the AGN sample, and the right panel displays the binned EW. }
    \label{fig:EW_grid}
\end{figure*}

Figure~\ref{fig:Lflux_grid} presents the comparison of ionizing luminosities, LOC-predicted line luminosities, and observed line luminosities for various emission lines in the \Mbh-\Redd\ plane. 
In the left panel, we compute the ionizing luminosities using the integral expression $L_{\rm ion} = \int_{\nu_{\rm product}}^{\nu_{\rm destruct}} L_\nu\,d\nu$ with the production and destruction edges: 47.9 and 64.5 eV for \CIV, 7.6 and 15.0 eV for \MgII, 24.6 and 54.4 eV for \HeIIUV, and 13.6 eV for \hbeta. The middle panel displays the line luminosities predicted by the LOC model using Eqn.~\ref{eq:LOC_integral} (see \S\ref{sec:photoionization}). The right panel presents the observed line luminosities for both local RM AGNs and the SDSS-RM sample. For local RM AGNs, \hbeta\ line luminosities are from \cite{Du&Wang2019}. The line luminosities for \halpha, \hbeta\ and \HeIIopt\ in the SDSS-RM sample are from \cite{Shen_etal_2019b}. 
There are offsets of $\sim 1$ dex in the observed \CIII, \HeIIopt, and \HeIIUV emission line luminosities to the LOC prediction, which may be a result of the underestimated EUV SED and the simplified assumptions of the  BLR cloud distribution. Despite these offsets, all panels in Figure~\ref{fig:Lflux_grid} exhibit a positive correlation in the $\Mbh-\Redd$ plane. Most importantly, this coherent trend between the LOC predictions and observations suggests that our observed emission line flux can be explained by the simple LOC model, and depends on the BH parameters and subsequently the ionizing SEDs. For even lower accretion rates ($\log{\dotm}<-1.5$), the computational limitations of {\tt qsosed} restrict us from evaluating the incident continuum and subsequently performing photoionization computations. 

We also compare theoretical and observed rest-frame equivalent widths (EWs) of several emission lines as functions of BH mass and Eddington ratio in Figure~\ref{fig:EW_grid}. In the left panel, we use the incident SED and LOC predicted line flux to predict the EW. For massive BHs ($\Mbh>10^8\,M_\odot$), more luminous (higher accretion-rate) BHs exhibit weaker emission line strength, particularly for UV lines, and vice versa for BHs with mass $\Mbh\leqslant10^8\,M_\odot$, as predicted by the LOC model. This pattern predicted by the LOC model indicates that the continuum and line luminosities change non-linearly with \Mbh\ and \Redd. This non-linear relationship suggests that the underlying SED shape may be more complex and varies significantly with changes in these parameters. The middle panel of Figure~\ref{fig:EW_grid} shows the observed EW values, and the right panel displays binned observed EW values. 
Although the predicted EWs in \CIII, \HeIIopt, and \HeIIUV are slightly higher than observed, the overall observed EW values remain generally consistent with the predictions from the LOC model. This consistency is expected as the observed continuum luminosities and line flux agree with the prediction in Figure~\ref{fig:Lconti_grid} and Figure~\ref{fig:Lflux_grid}, respectively. However, due to sample selection biases for the RM AGNs (i.e., missing low-mass, low-accretion-rate objects), it is difficult to explore the EW in the lower-left corner of the \Mbh-\Redd\ space, thus unable to demonstrate the same diagonal trend in the LOC model prediction.



\subsection{Comparisons of composite UV spectra}\label{sec:composite_spec}

\begin{figure*}
\centering
    \includegraphics[width=\linewidth]{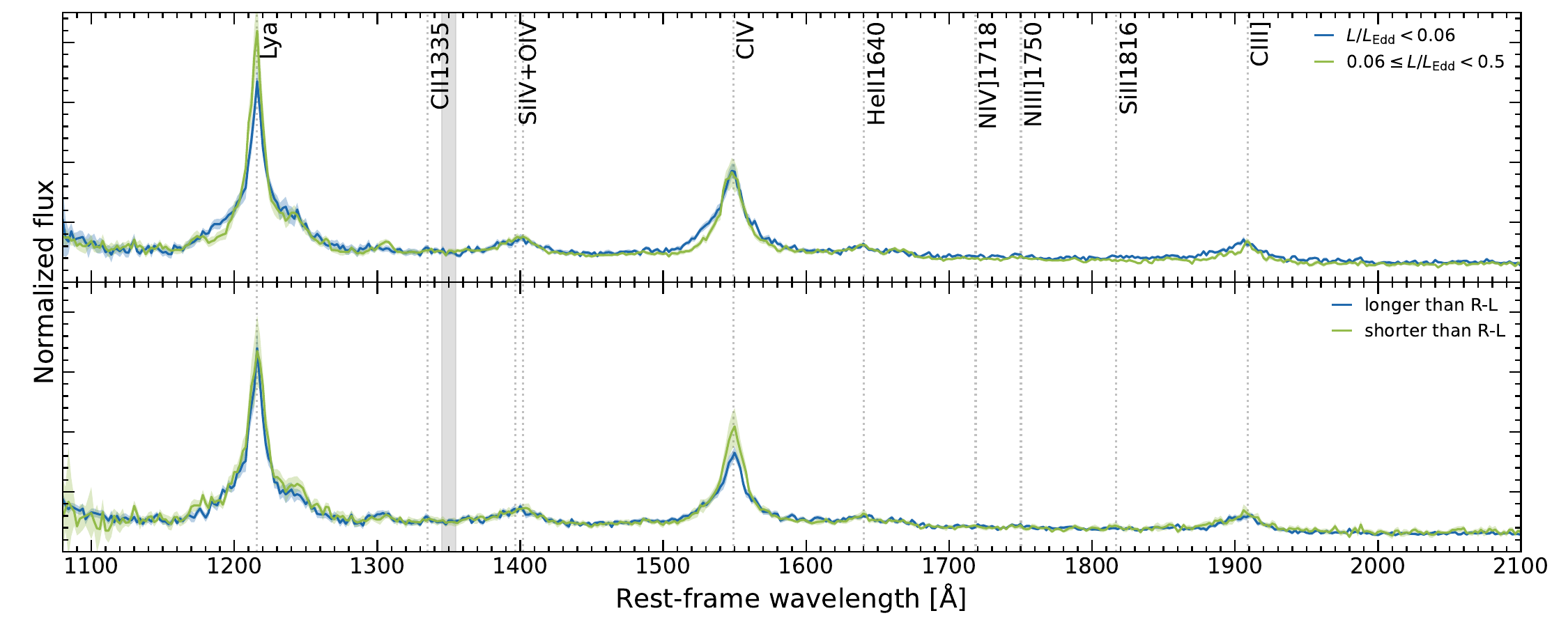}
    \caption{Median composite spectra for different accretion rate samples (top) and different \hbeta\ lag samples (bottom) with roughly matched optical luminosity $43 \leqslant \log(\L5100) \leqslant 45$, normalized at 1350 ${\rm\AA}$ (grey shaded area).}
    \label{fig:composite_spec}
\end{figure*}

Since observed (and measured) properties in individual objects can suffer from significant measurement uncertainties, we also use the higher S/N composite spectra to compare with theoretical predictions. Figure~\ref{fig:composite_spec} presents the composite rest-frame UV  spectra for RM AGNs within a matched optical luminosity range ($43 \leqslant \log(\L5100) \leqslant 45$) but with different \Redd (upper panel) and \hbeta\ lags (lower panel). The fluxes are normalized at rest-frame wavelength 1350 ${\rm\AA}$ to better illustrate the UV line strength. 
In the upper panel, subsets with higher \Redd\ values (green) exhibit stronger \lya\ line strengths compared to the lower accretion group (blue). According to the SED example in Figure~\ref{fig:SED_fix5100} and our photoionization results in Figure~\ref{fig:EW_grid}, at a matched optical luminosity, BHs with higher accretion rates possess more ionizing energy for both hydrogen and \CIV, resulting in stronger emission lines. This trend is evident in our composite spectra, where higher accretion rate samples tend to show stronger emission lines.  This agreement demonstrates that our photoionization models can accurately generate line strengths based on the given BH parameters.

Likewise, in the lower panel of Figure~\ref{fig:composite_spec}, we divide our sample into two subsets based on their \hbeta\ time lag relative to the canonical $R-L$ relation. The subset with lags shorter than the $R-L$ relation (green) has stronger \CIV\ line strength compared to the longer lag sublet (blue); while these two spectra appear to have similar \lya\ line strength. Given that \CIV\ is a collisionally excited line with an excitation energy of 8 eV, the shorter-lag subset suggests that it may have an additional ionizing flux at FUV compared to the longer-lag subset, with the former corresponding to the higher-accreting systems in Figure~\ref{fig:SED_fix5100}. However, it appear to contradict our BLR size prediction in Figure~\ref{fig:theory_R_L_grid} that higher-accreting objects should have more ionizing energy and thus longer lags. This discrepancy may be due to the constant parameters of our BLR size prediction, which will be discussed further in \S\ref{sec:R-L_dispersion}.


\section{Discussion}\label{sec:discussion}

\subsection{Line ratios as a proxy for SED shape}\label{sec:line_ratio_SEDshape}

\begin{figure*}
    \includegraphics[width=\linewidth]{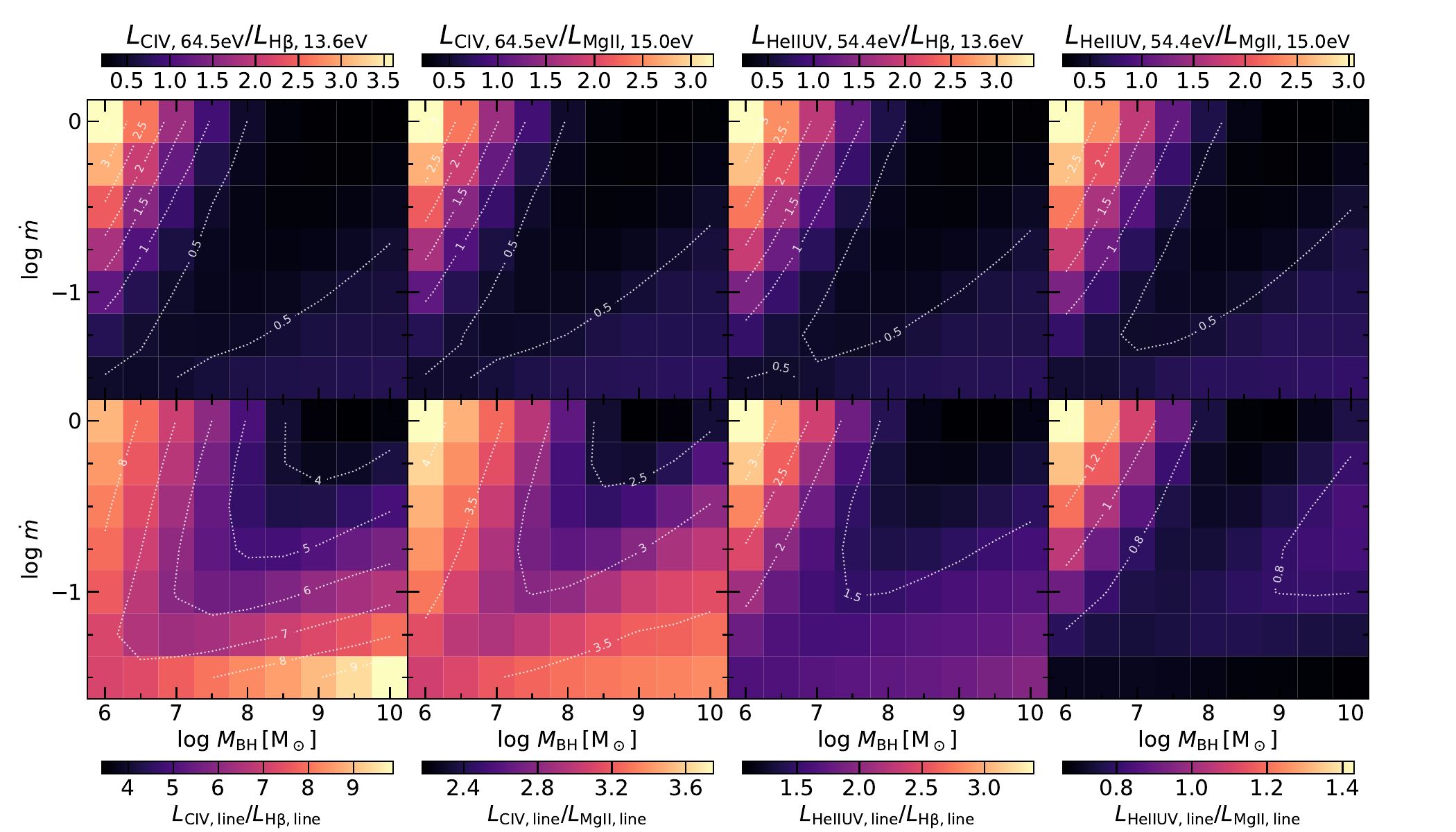}
    \caption{Contour plots showing the dependence on the mass and accretion rate for monochromatic luminosity ratio at certain energies (upper panel) and various predicted line ratios (lower panel).}
    \label{fig:Lev_vs_Lline_ratio}
\end{figure*}

Our LOC model photoionization calculations have shown qualitative agreement with the observed line properties (line strengths, BLR distances, etc.). Since lines with different ionization potentials depend on different parts of the SED, it is desirable to infer the underlying ionizing SED (not directly observable) from observables, such as the flux ratios between low- and high-ionization lines. 

Figure~\ref{fig:Lev_vs_Lline_ratio} displays several groups of continuum luminosity ratios and line flux ratios as functions of \Mbh\ and \Redd, predicted from photoionization calculations. In the top panel of Figure~\ref{fig:Lev_vs_Lline_ratio}, the monochromatic luminosities at the ionizing energies of \CIV (64.5 eV), \MgII (15.0 eV), \hbeta (13.6 eV) and \HeIIUV (54.4 eV) are computed from the grid of SEDs and then their line ratios are placed in the \Mbh-\Redd\ plane. The bottom panel shows the corresponding line luminosity ratios of the same lines as in the top panel. The line luminosity ratio is computed using {\tt CLOUDY} and the LOC model, as explained in \S\ref{sec:photoionization}. Both monochromatic luminosity ratios and line luminosity ratios demonstrate complicated dependence on the mass and accretion rate.
Given that \CIV\ and \MgII\ are collisionally excited lines with excitation energies of 8 eV and 4.4 eV, while \hbeta\ and \HeIIUV\ are recombination-dominated lines, the line ratios involving \CIV\ may not accurately trace the SED shape; in contrast, \HeIIUV\ is more effective in tracing the SED shape in Figure~\ref{fig:Lev_vs_Lline_ratio}.

\begin{figure}
    \centering
    \includegraphics[width=\linewidth]{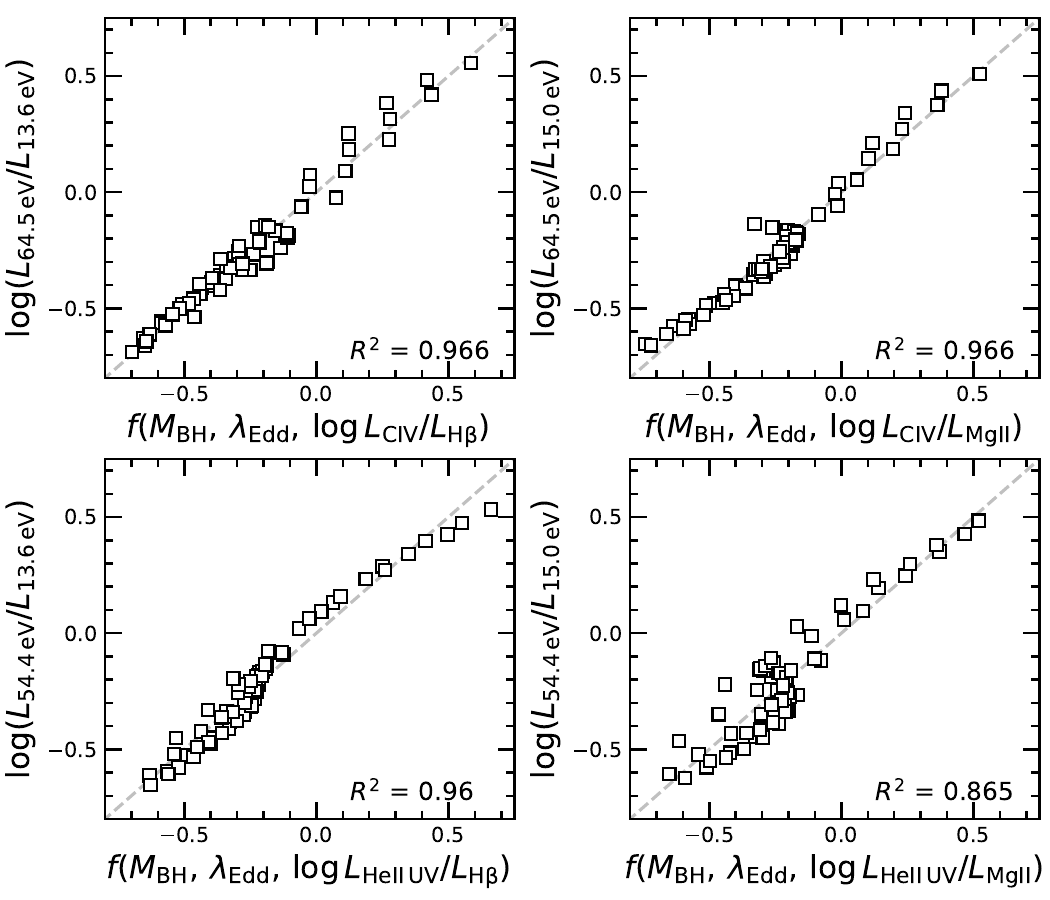}
    \caption{Scaling relations of monochromatic luminosity ratios (i.e. SED shape) and observables (\Mbh, \Redd\ and lines flux ratio) for the theoretical models, with the coefficients of determination $R^2$ on the upper left corners.}
    \label{fig:lfr_prob_regression}
\end{figure}

Given the strong correlations between predicted continuum-luminosity ratios and line-flux ratios in Figure~\ref{fig:Lev_vs_Lline_ratio}, we use the BH mass \Mbh, accretion rate \Redd, and line flux ratios in our theoretical computation models to provide empirical relations that map the underlying ionizing SED shape from $\sim 10$ to $\sim 60$ eV. The concept of using UV emission line strengths to probe the ionizing SED has been previously applied to the EW of \HeIIUV as tracers of the EUV continuum \citep[e.g.,][]{Mathews&Ferland1987, Korista&Goad_2000, Leighly2004, Temple_etal_2023}. However, this approach does not consider the possibility that the ionizing continuum that the BLR gas receives is not the same as what observers see, thus \HeIIUV\ EW is not tracing the real ionizing continuum shape but the ratio between ionizing energy for \HeIIUV\ and the continuum that the observer receives \citep{Timlin_etal_2021}. To address this issue, we use the flux ratio of different UV/optical line pairs to directly map the underlying SED shape, which involves only the ionizing continuum received by the BLR gas.

To demonstrate the ability of observables to trace the SED shape with empirical relations, we use linear regression to fit the theoretical results with the following relation:
\begin{equation}
\begin{split}\label{eqn:LE_ratio}
    \log(L_{E_1}/L_{E_2}) = &\theta_0 + \theta_1 \log{M_{\rm BH}} + \theta_2 \log{\Redd} \\
    &+ \theta_3 \log(L_{\rm line1}/L_{\rm line2}),
\end{split}
\end{equation}
where $L_E$ is the monochromatic luminosity at energy $E$, and $E_1$ and $E_2$ are the ionizing energy of line1 and line2, respectively. 
The best-fit parameters are listed in Table~\ref{tab:para_LE_ratio}.
The scaling relations between the observables and the ionization continuum shape are shown in Figure~\ref{fig:lfr_prob_regression}. The coefficients ($R^2= 1 - {\sum_{i=1}^{n} (y_i - \hat{y}_i)^2}/{\sum_{i=1}^{n} (y_i - \bar{y})^2}
$) of the four pairs of lines suggest that the line ratios can trace the ionizing continuum shape at $\sim 10$ eV fairly well, especially for \CIV/\MgII.

\begin{table}
\caption{Best-fit parameters of line excitation energy ratios $\log(L_{E_1}/L_{E_2})$ defined in Eqn.~\ref{eqn:LE_ratio}}\label{tab:para_LE_ratio}
\centering
\scalebox{0.8}{
\begin{tabular}{lcccc}
\hline\hline
Line ratios & $\theta_0$ & $\theta_1$ & $\theta_2$ & $\theta_3$ \\
\hline
CIV/\hbeta & -0.894 & -0.169 & 0.567 & 3.163 \\
\CIV/\MgII & -1.602 & -0.109 & 0.335 & 5.362 \\
\HeIIUV/\hbeta\ & -0.327 & -0.045 & 0.106 & 2.535 \\
\HeIIUV/\MgII & -0.451 & 0.040 & -0.257 & 3.163 \\
\hline\hline
\end{tabular}
}
\end{table}


Multi-wavelength observations probing additional emission lines with different ionizing energies will allow mapping the ionizing continuum to an even wider range. For example, the line ratios of infrared coronal lines (CLs) from different elements and different excitation states have proven to be related to the intrinsic BH properties (\Mbh\ and $\dot{m}$), chemical abundance, and ionization parameters \citep{Ferguson_etal_1997}. Most importantly, CLs have the capability to map the ionizing continuum to energy levels exceeding 100 eV, making these data more powerful than UV data alone. Photoionization computations in \cite{Cann_etal_2018} demonstrate that several pairs of CL ratios might trace intermediate-mass BHs in dwarf galaxies, as the harder SEDs in IMBHs can excite stronger high-ionization lines. With the advent of the \textit{James Webb Space Telescope} (JWST), infrared spectroscopic observations covering a wealth of IR coronal lines will provide a viable approach to studying the connections between the ionizing continuum and CL emission, and to understanding the physical driver of the diversity in quasar line strength.

\subsection{Dispersion in the \hbeta\ $R-L$ relation}\label{sec:R-L_dispersion}

\begin{figure}
    \centering
    \includegraphics[width=\linewidth]{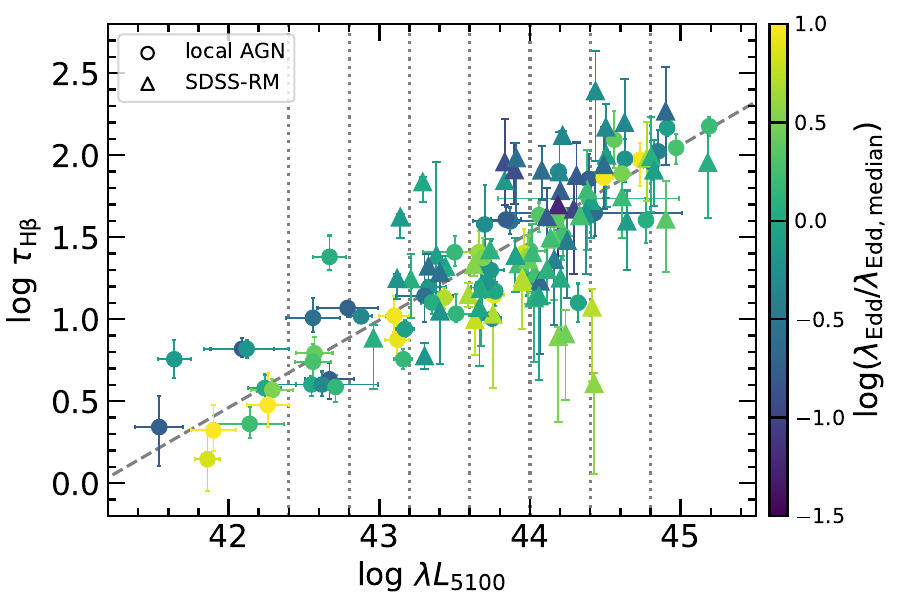}
    \caption{The $R-L$ relation: observed \hbeta\ lag against the optical luminosity for RM quasars with $\Redd<0.5$, color-coded based on the relative $\log\Redd$ values calculated as the ratio of \Redd\ relative to the median \Redd\ in each $\L5100$ bin (suggested by grey dotted lines).}
    \label{fig:R-L_Ledd_binned}
\end{figure}

In \S\ref{sec:predict_R-L} we predicted the \hbeta\ time lags with constant ionization parameter and compared them with the observed lags. Since the BLR is sensitive to the entire ionizing flux emitted by the central source, varying \Mbh\ and \Redd\ lead to different ionizing SEDs that are responsible for different BLR sizes as well as different line-ionizing fluxes for different line-emitting ions in the BLR clouds. The parameters (\Mbh\ and \Redd) are degenerate at fixed optical luminosity in the $R-L$ relation, as shown in Figure~\ref{fig:SED_fix5100} and Figure~\ref{fig:Lconti_grid}. It is therefore natural to expect an intrinsic dispersion in time lags at fixed optical luminosity caused by different accretion parameters and subsequently the underlying SED. So far this model has been the most favored interpretation in recent work on this topic \citep{Du&Wang2019, Fonseca_etal_2020, Shen_etal_2024, Czerny_etal_2019, Martinez-Aldama_etal_2019,Dalla_etal_2020}. Moreover, smearing our theoretical predictions with observed uncertainties demonstrates that uncertainties in the measured time lag along with luminosity variability can introduce additional scatter in the $R-L$ relation.

The scale of the BLR can also be affected by other accretion parameters that were not included in this preliminary study. For example, in the high-accretion regime (approaching the Eddington limit), the accretion disk structure may significantly depart from the SSD model \citep{Shakura&Sunyaev1973}, and become a slim disk \citep{Abramowicz1988} where the disk scale height is comparable to the radius. The slim disk geometry will produce anisotropic radiation fields due to the self-shadowing effect, which could reduce the ionizing flux and soften the SED that the BLR cloud receives. In particular, systematically shorter \hbeta\ time lags in high-accretion quasars have been reported by the SEAMBH collaboration \citep{Du2014_SEAMBH, Du2015_SEAMBH, Du2018_SEAMBH, Hu2021_SEAMBH}. Due to considerable theoretical uncertainties in the slim disk SED predictions and the relative paucity of UV spectroscopy for RM AGNs accreting in this regime, we defer a more complete investigation on AGNs with slim-disk accretion to a future paper.

It is important to note, however, that our investigations suggest that in the low-to-moderate-accretion thin-disk regime and at fixed optical luminosity, AGNs with larger BH masses or lower accretion rates tend to have shorter lags. This prediction is in tension with the observed trend. Figure~\ref{fig:R-L_Ledd_binned} presents the observed $R-L$ relation for our RM AGN sample, color-coded by the relative Eddington ratio at each bin. A decreasing trend of lags with $L/L_{\rm Edd}$ at fixed optical luminosity is observed even for the SSD regime studied in this work. This discrepancy suggests that certain simple assumptions in our photoionization modeling, e.g., a universal gas density, might be incorrect. For example, if $n_{\rm H}$ increases with $L/L_{\rm Edd}$, and given that $R_{\rm BLR} \propto n_{\rm H}^{-1/2}$ (Egn.~\ref{eq:R_BLR}), we expect a more compact BLR size from the LOC model, which would be more consistent with the observed trend in the $R-L$ plane. Additionally, as $n_{\rm H}$ increases, broad-line emissivity would also increase, resulting in higher predicted broad-line fluxes for higher Eddington ratios, again consistent with observations. 



\subsection{Predicted hydrogen-ionizing photon flux}  \label{sec:predict_QH}

\begin{figure}
    \centering
    \includegraphics[width=\linewidth]{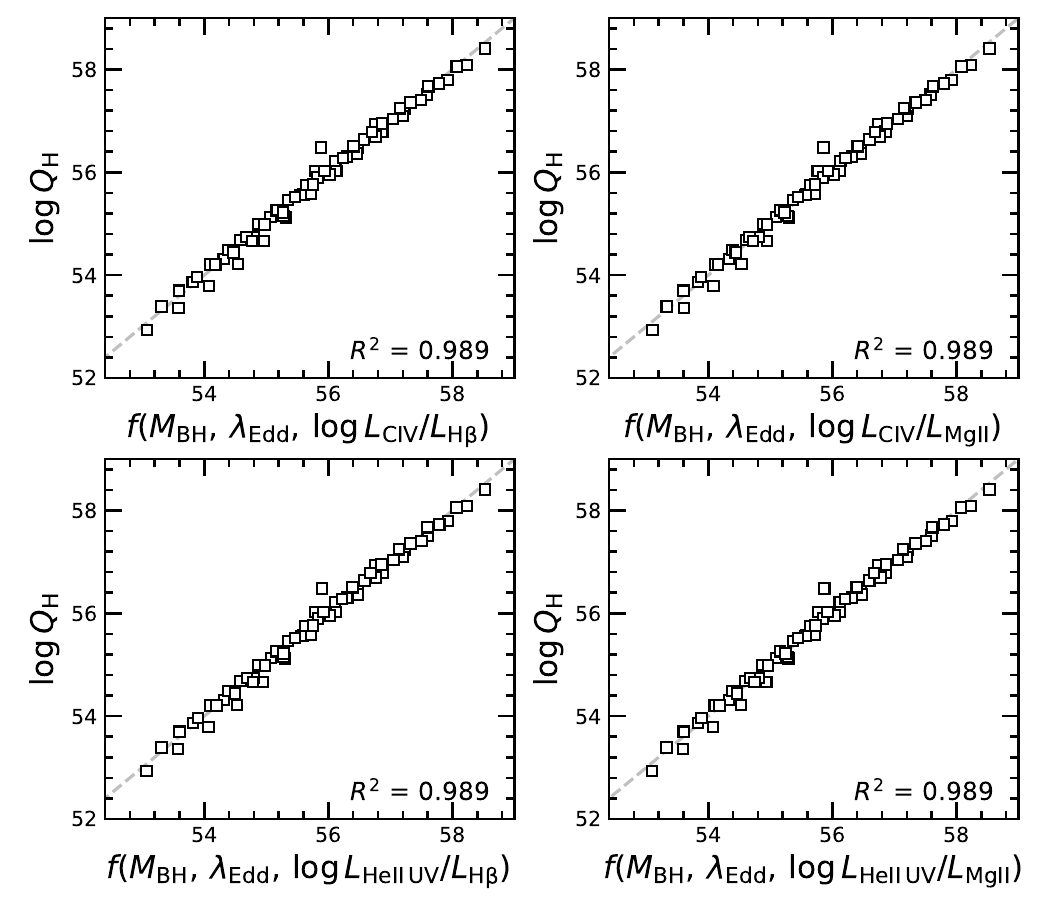}
    \caption{Linear regression scaling between the ionizing photon number $Q({\rm H})$ and the observables for theoretical measurements.}
    \label{fig:QH_regression}
\end{figure}

\begin{figure}
    \centering
    \includegraphics[width=\linewidth]{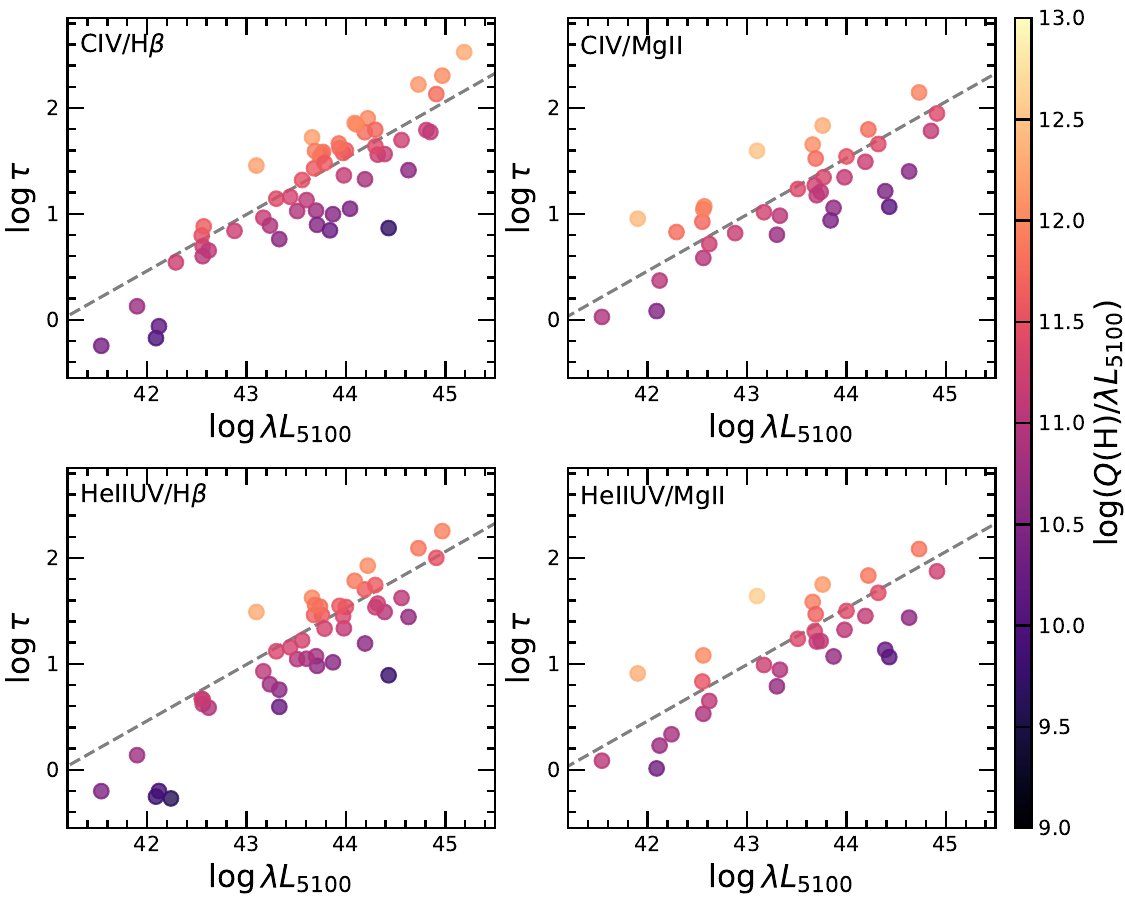} 
    \caption{Predicted \hbeta\ time lag following Eqn.~(\ref{eq:R_BLR}) versus the observed optical luminosity, color-coded by $Q({\rm H})/\L5100$, where the ionizing photon flux $Q({\rm H})$ is predicted from observed parameters (Eqn.~\ref{eqn:Qh}). 
    }
    \label{fig:R-L_QH}
\end{figure}


\begin{table}
\caption{Best-fit parameters of the hydrogen-ionizing photon flux $\log Q({\rm H})$ in Eqn.~\ref{eqn:Qh}}\label{tab:para_QH}
\centering
\scalebox{0.8}{
\begin{tabular}{lcccc}
\hline\hline
Line ratios & $\theta_0$ & $\theta_1$ & $\theta_2$ & $\theta_3$ \\
\hline
CIV/\hbeta & 53.035 & 0.475 & 1.636 & 0.358 \\
\CIV/\MgII & 53.140 & 0.479 & 1.581 & 0.200 \\
\HeIIUV/\hbeta\ & 53.057 & 0.492 & 1.583 & 0.342 \\
\HeIIUV/\MgII & 53.119 & 0.493 & 1.546 & 0.282 \\
\hline\hline
\end{tabular}
}
\end{table}

In our theoretical analysis, the time lag of \hbeta is modeled as a linear function of the hydrogen-ionizing photon flux $Q({\rm H})$, as expressed by Eqn.~\ref{eq:R_BLR}. Because the observed ionizing continuum may not be the same as the BLR receives, the flux ratios of emission lines from species with different ionization potentials are the most promising way to constrain the ionizing photon flux that directly determines the BLR size \citep{Negrete_etal_2012, Ferland_etal_2020, BuendiaR_etal_2023}. As discussed in \S\ref{sec:line_ratio_SEDshape}, the line-flux ratios, coupled with the mass and Eddington ratio, have demonstrated their potential in deriving the SED shape. Therefore, we adopt a similar approach to predict the $Q({\rm H})$ for observed quasars through the following equation:
\begin{equation}\label{eqn:Qh}
\begin{split}
    \log Q({\rm H}) = &\theta_0 + \theta_1 \log{M_{\rm BH}} + \theta_2 \log{\Redd} \\
    &+ \theta_3 \log(L_{\rm line1}/L_{\rm line2}),
\end{split}
\end{equation}
We first fit a linear regression relation on the parameter grid derived from theoretical models, and the best-fit parameters are listed in Table \ref{tab:para_QH}. The linear regression scaling is shown in Figure~\ref{fig:QH_regression} for the four line pairs, where the ionizing photon flux $Q({\rm H})$ could be well estimated using such observables (\Mbh, \Redd, and line flux ratios). 

Subsequently, we apply the regression parameters to the observed RM quasars. Figure~\ref{fig:R-L_QH} presents color-coded $R-L$ relation for low-accretion AGNs with their predicted $Q({\rm H})/\L5100$ value. Since our assumption in \S\ref{sec:predict_R-L} implies that $R_{\rm BLR}\sim Q^{0.5}$, suggesting that at any given $\L5100$, \hbeta\ lags would follow a vertical increase as $Q({\rm H})/\L5100$, such a vertical trend can be seen in Figure~\ref{fig:R-L_QH}. However, in Figure~\ref{fig:R-L_Ledd_binned}, \hbeta\ lags show decreasing trend as relative accretion rates increase, suggesting the opposite trend to Figure~\ref{fig:R-L_QH}. This discrepancy is possibly due to over-simplified assumptions (e.g., the constant hydrogen density) in our BLR size prediction (see \S\ref{sec:R-L_dispersion}).



\subsection{Uncertainties and limitations}\label{sec:uncer}

The primary goal of this paper is to explore the first-order effects of varying the BH mass and accretion rate on the BLR distance and the diverse line strengths within the LOC model framework. We emphasize that we have not attempted to construct a model finely tuned to match the properties of any specific AGN. This general approach quantitatively reproduces the behavior of the BLR distance and line luminosity relative to BH mass and accretion rate given universal ionization parameters. However, this photoionization modeling is carried out for low-accretion quasars in the SSD regime with fixed inclination $\cos{i}=0.5$ and for non-spinning BHs only. The impact of BH spin on the BLR size is pronounced, which would affect the shape of the ionizing continuum, especially in the UV band \citep{Kubota&Done2018, Wang_etal_2014a, Reynolds2021_BHspin}. The inclusion of BH spin in our analysis would introduce additional complexity, potentially leading to a greater dispersion in the $R-L$ relation \citep{Czerny_etal_2019} and producing degenerating results with BH mass and accretion rate.

In our SED modeling, because of the imposed accretion rate limits of  $-1.65<\log\dotm<0.39$ in {\tt qsosed}, this work is limited to mostly the SSD-dominated accretion regime and only models $\log\dotm\leq0$. Compared to the original {\tt agnsed}, the simplified model {\tt qsosed} fixes the electron temperature and spectral index at typical values for the warm and hot Comptonization components as follows: $kT_{\rm e, hot} = 100\, \keV$, $kT_{\rm e, warm} = 0.2\, \keV$, $\Gamma_{\rm warm} = 2.5$, $r_{\rm warm} = 2r_{\rm hot}$, $r_{\rm out} = r_{\rm sg}$, $h_{\rm max} = 100$, with the power dissipated by the hot Comptonization region fixed at $0.02L_{\rm Edd}$ \citep{Kubota&Done2018}. 
By comparing the stacked AGN SEDs from SOUX \citep{Kynoch_etal2023} with {\tt qsosed} predictions, \cite{Mitchell_etal2023} found that while the {\tt qsosed} model matches well with the observed SEDs for intermediate-mass AGNs ($\log\Mbh = 7.5 - 9.0$), it underpredicts the X-ray flux at low-\Mbh\ and shows discrepancies in the optical/UV continuum for high-\Mbh\ AGNs. 
Assuming that the X-ray emission is generated from the hot Comptonization region and the EUV emission originates from the warm Comptonization region \citep{Petrucci_etal2018, Kubota&Done2018}, the absence of a predictive model with solid theoretical basis to untangle the poorly understood warm Comptonization component and its relation to the hot Comptonization coronal properties still limits the comparison between observations and theoretical analysis, and thereby our understanding of AGN SEDs.
However, the demonstrated correlations between the X-ray, EUV, and UV regions \citep[e.g.,][]{Green1996,Steffen_etal_2006, Just_etal_2007,Lusso&Risaliti2016,Timlin_etal_2020a,Timlin_etal_2021} may provide crucial insights that could uncover the nature of the hot/warm corona in quasars.
Despite that this simplified {\tt qsosed} model may provide limited flexibility in SED shapes, our LOC photoionization computations indicate that this straightforward model can successfully reproduce the line strengths as observed for various BH mass and accretion rates. 
As discussed in \S\ref{sec:R-L_dispersion}, as the accretion rate substantially increases beyond the SSD-dominated regime, the accretion disk solution may be better described by the slim disk model with a puffed-up inner disc that shadows outer regions, which will potentially soften the SED received by the BLR clouds \citep{Abramowicz1988, Wang2014}. Due to the complexities involved in modeling such a scenario and the uncertainties in slim disk SED prediction, we postpone a comprehensive investigation of these high-accretion AGNs to future work.

While we demonstrate the potential of using line ratios to probe the SED shape and the ionizing flux, the predicted line ratios may deviate from observed line ratios. This discrepancy can arise from factors such as differences in metallicity and assumptions regarding various ionizing conditions \citep{Panda_etal_2019, Korista&Goad2019}. While mechanisms other than BH mass and accretion rate will have some effects on the BLR size and observed line strength, we have demonstrated that variations in the \Mbh-\Redd\ have successfully reproduced the lags and emission-line properties over many orders of magnitude in luminosity and BH mass. Nevertheless, discrepancies remain in the detailed predictions of the lag dispersion at fixed optical luminosity, as discussed in \S\ref{sec:R-L_dispersion}.

Finally, there is the possibility that the BLR gas sees a different ionizing continuum than the observer, due to a potential inner shielding gas component that blocks some of the highest energy photons \citep[e.g.,][]{Wu_etal2011_WLQ, Wu_etal2012_WLQ, Luo_etal2015_WLQ, Plotkin_etal2015, Ni_etal_2018, Ni_etal_2022}. In this work, we do not see much evidence for this shielding gas, as the predicted line strengths are consistent with the BLR seeing all the continuum flux. However, the covering fraction \citep{Ferland_etal_2020} and the impact of this shielding gas may become more prominent for even higher Eddington ratios, i.e., in the slim disk regime. We plan to investigate this possibility in future work.

\section{Conclusions}\label{sec:con}


In this paper, we have performed {\tt CLOUDY} photoionization calculations with the empirical LOC model and typical parameters to predict the broad-line emission as functions of accretion parameters, BH mass, and accretion rate, using the latest AGN SED models \citep{Kubota&Done2018} as the input continuum. We compare our theoretical results with observed AGN properties for a sample of 70 low-redshift AGNs with reverberation mapping measurements that are in the sub-Eddington accretion regime. Our main conclusions are as follows:

\begin{enumerate}
    \item {Assuming universal values of hydrogen density ($\log n({\rm H})=12$) and ionization parameter ($\log U_{\rm H}=-2$), the photoionization models reproduce the observed global relation between the Balmer broad-line lag and optical luminosity at rest-frame 5100\,\AA\ (\S\ref{sec:predict_R-L}). However, at fixed $L_{\rm 5100}$, the model predicts longer lags for higher Eddington ratios, opposite to the observed trend. These results may imply that the gas density increases with the accretion rate to reduce the average size of the BLR. (\S\ref{sec:R-L_dispersion}) }
    
    \item{The photonization models produce increasing optical and UV broad-line strengths with accretion rate at fixed BH mass or optical luminosity, which are consistent with observations. Overall, the distributions of different broad-line fluxes in the \Mbh-\Redd\ plane are consistent between observations and theoretical predictions (\S\ref{sec:obs_line_strength}) }
    
    
    
    \item{Based on these comparisons, we provide empirical scaling relations (Eqn.~\ref{eqn:LE_ratio} and \ref{eqn:Qh}) that utilize the observed BH mass, $L/L_{\rm Edd}$, and line flux ratios to constrain the ionizing continuum flux, which is not directly observable. (\S\ref{sec:predict_QH}) }

\end{enumerate}

The overall consistency between observations of AGN broad-line properties and photoionization modeling suggests that such analyses are important to fully decipher the accretion parameters and the resulting broad-line emission in broad-line AGNs. The optical and UV broad emission lines only sample a limited portion of the ionizing continuum, and future observations with high-ionization coronal lines (with ionizing energy at $>150\,{\rm eV}$) can lead to better constraints on the overall continuum SED. Further refinements of the LOC model and AGN continuum models, as well as extending to higher accretion rates better described by the slim disk model, will be crucial to fully understanding the dispersion of lags at fixed optical luminosity, an important step to designing more accurate single-epoch BH mass recipes using the $R-L$ relation. 


\begin{acknowledgments}
Q.W. and Y.S. acknowledge partial support from NSF grant AST-2009947. Based on observations with the NASA/ESA Hubble Space Telescope obtained from the Data Archive at the Space Telescope Science Institute, which is operated by the Association of Universities for Research in Astronomy, Incorporated, under NASA contract NAS5-26555. Support for programs HST-GO-16171 and HST-GO-17433 was provided through a grant from the STScI under NASA contract NAS5-26555. 
HXG is supported by the National Key R\&D Program of China No. 2022YFF0503402, 2023YFA1607903.
LCH was supported by the National Science Foundation of China (11991052, 12233001), the National Key R\&D Program of China (2022YFF0503401), and the China Manned Space Project (CMS-CSST-2021-A04, CMS-CSST-2021-A06). 
J.I.L. is supported by the Eric and Wendy Schmidt AI in Science Postdoctoral Fellowship, a Schmidt Futures program. 

\end{acknowledgments}

\bibliography{refs}{}
\bibliographystyle{aasjournal}


\end{document}